\documentclass[12pt,twoside, a4paper]{article}
\def\pd{\partial}
\def\mc{\mathcal}

\usepackage[dvips]{graphicx}
\usepackage{amssymb}
\usepackage{amssymb,amsmath}
\usepackage{graphicx}
\usepackage{dsfont}
\usepackage{caption}
\usepackage{subcaption}
\usepackage{mathtools}
\usepackage{verbatim}
\usepackage{graphicx}
\usepackage{multirow}
\usepackage[outline]{contour}
\usepackage{xcolor,colortbl}
\usepackage{pdflscape}
\input{epsf.sty}  
\begin{document}
\begin{center}
\LARGE{\textbf{Holographic RG flows and Janus solutions from matter-coupled $N=4$ gauged supergravity}}
\end{center}
\begin{center}
\large{\textbf{Parinya Karndumri}}
\end{center}
\begin{center}
String Theory and Supergravity Group, Department
of Physics, Faculty of Science, Chulalongkorn University, 254 Phayathai Road, Pathumwan, Bangkok 10330, Thailand
\end{center}
E-mail: parinya.ka@hotmail.com \vspace{1 cm}\\
\begin{abstract}
We study an $SO(2)\times SO(2)\times SO(2)\times SO(2)$ truncation of four-dimensional $N=4$ gauged supergravity coupled to six vector multiplets with $SO(4)\times SO(4)$ gauge group and find a new class of holographic RG flows and supersymmetric Janus solutions. In this truncation, there is a unique $N=4$ supersymmetric $AdS_4$ vacuum dual to an $N=4$ SCFT in three dimensions. In the presence of the axion, the RG flows generally preserve $N=2$ supersymmetry while the supersymmetry is enhanced to $N=4$ for vanishing axion. We find solutions interpolating between the $AdS_4$ vacuum and singular geometries with different residual symmetries. We also show that all the singularities are physically acceptable within the framework of four-dimensional gauged supergravity. Accordingly, the solutions are holographically dual to RG flows from the $N=4$ SCFT to a number of non-conformal phases in the IR. We also find $N=4$ and $N=2$ Janus solutions with $SO(4)\times SO(4)$ and $SO(2)\times SO(2)\times SO(3)\times SO(2)$ symmetries, respectively. The former is obtained from a truncation of all scalars from vector multiplets and can be regarded as a solution of pure $N=4$ gauged supergravity. On the other hand, the latter is a genuine solution of the full matter-coupled theory. These solutions describe conformal interfaces in the $N=4$ SCFT with $N=(4,0)$ and $N=(2,0)$ supersymmetries.    
\end{abstract}
\newpage

\section{Introduction}
Over the past twenty years, solutions of gauged supergravities in different space-time dimensions have provided a large number of useful holographic descriptions for strongly coupled dual field theories. Among various types of these solutions, holographic RG flows and Janus solutions are of particular interest since the original proposal of the AdS/CFT correspondence \cite{maldacena,Gubser_AdS_CFT,Witten_AdS_CFT}. These explain deformations of the dual conformal field theories (CFT) and the presence of conformal interfaces or defects within the parent CFTs. The most useful results along this direction arise when there is some amount of unbroken supersymmetry in which many properties at strong coupling are controllable.   
\\
\indent Although there are AdS/CFT dualities in many space-time dimensions, AdS$_4$/CFT$_3$ correspondence attracts much attention due to the relevance in describing world-volume dynamics of M2-branes, the fundamental degrees of freedom in M-theory. In this case, four-dimensional gauged supergravities are very useful in obtaining holographic solutions of interest which in some cases can be uplifted to ten or eleven dimensions. A number of holographic RG flows and Janus configurations have previously been found in gauged supergravities with various gauge groups and different numbers of supersymmetries \cite{Warner_membrane_flow,Warner_M_F_theory_flow,Warner_higher_Dflow,Flow_in_N8_4D,4D_G2_flow,
Warner_M2_flow,Guarino_BPS_DW,Elec_mag_flows,Yi_4D_flow,N3_SU2_SU3,N3_4D_gauging,tri-sasakian-flow,orbifold_flow,4D_N4_flows,N5_flow,N3_Janus,Minwoo_4DN8_Janus,Kim_Janus,omega_DW_1,omega_DW_2,
N6_flow,N8_omega_Janus}, for similar solutions with other space-time dimensions, see \cite{7Dflow,6Dflow,6Dflow1,6Dflow2,GPPZ,FGPP,5Dflow_Davide,5Dflow_bobev,5D_flow_HL,3D_Janus_de_Boer,
3D_Janus_Bachas,3D_Janus_Bak,half_BPS_AdS3_S3_ICFT,exact_half_BPS_string,multi_face_Janus,4D_Janus_from_11D,
6D_Janus,3D_Janus} for an incomplete list.  
\\
\indent In this paper, we will give a new class of holographic RG flows and supersymmetric Janus solutions within matter-coupled $N=4$ gauged supergravity with $SO(4)\times SO(4)$ gauge group. This gauged supergravity can be constructed by coupling pure $N=4$ gauged supergravity to $n\geq 6$ vector multiplets. We will consider the simplest case of $n=6$ here. In this gauged supergravity, a number of supersymmetric $AdS_4$ vacua and RG flows between them have recently appeared in  \cite{4D_N4_flows}. In that case, the truncation to $SO(3)\times SO(3)$ singlet scalars have been considered and solutions with only the dilaton and the axion in the gravity multiplet or solutions with vanishing axion are consistent with supersymmetry. In addition, all solutions given in \cite{4D_N4_flows} preserve $N=4$ supersymmetry. In the present work, we will find more general solutions namely solutions with both the axion and vector multiplet scalars non-vanishing and breaking the original $N=4$ supersymmetry to $N=2$. 
\\
\indent We will also find a new class of supersymmetric Janus solutions in $N=4$ gauged supergravity which has not been considered in \cite{4D_N4_flows}. To the best of the author's knowledge, only Janus solutions found in \cite{tri-sasakian-flow} and \cite{orbifold_flow} are obtained within the framework of $N=4$ gauged supergravity. All the solutions given in \cite{tri-sasakian-flow} are the so-called singular Janus in the sense that they interpolate between non-conformal phases rather than between conformal phases, so these are holographically expected to describe conformal interfaces within non-conformal field theories in three dimensions. On the other hand, regular Janus solutions given in \cite{orbifold_flow} only involve the dilaton and axion from the gravity multiplet. There is also a singular Janus solution found in \cite{orbifold_flow} that involve only scalars from vector multiplets. Therefore, we believe that the $N=2$ Janus solution obtained in this paper is the first regular Janus solution in the context of matter-coupled $N=4$ gauged supergravity that involves scalars from both gravity and vector multiplets.    
\\
\indent The paper is organized as follows. In section \ref{N4_SUGRA},
we review the general structure of $N=4$ gauged supergravity coupled to vector multiplets in the embedding tensor formalism and at the end focus on the case of six vector multiplets with $SO(4)\times SO(4)$ gauge group. The truncation to $SO(2)\times SO(2)\times SO(2)\times SO(2)$ singlet scalars, $AdS_4$ vacua and general structure of relevant BPS equations in both RG flow and Janus solutions are also given in detail. This sets up the framework for finding solutions in subsequent sections. In sections \ref{RG_flow} and \ref{Janus}, a number of $N=4$ and $N=2$ holographic RG flow and Janus solutions are given, respectively. We end the paper by giving some conclusions and comments on the results in section \ref{conclusion}.

\section{Matter-coupled $N=4$ gauged supergravity}\label{N4_SUGRA} 
In four dimensions, $N=4$ supersymmetry allows two types of supermultiplets, the gravity and vector multiplets. The former contains the following field content
\begin{equation}
(e^{\hat{\mu}}_\mu,\psi^i_\mu,A_\mu^m,\chi^i,\tau)
\end{equation}
which are given by the graviton $e^{\hat{\mu}}_\mu$, four gravitini $\psi^i_\mu$, six vectors
$A_\mu^m$, four spin-$\frac{1}{2}$ fields $\chi^i$ and one complex
scalar $\tau$. The scalar consists of the dilaton $\phi$ and the axion $\chi$ and can be parametrized by $SL(2,\mathbb{R})/SO(2)$ coset. Indices $\mu,\nu,\ldots =0,1,2,3$ and $\hat{\mu},\hat{\nu},\ldots=0,1,2,3$ are respectively space-time and tangent space indices while $m,n=1,\ldots, 6$ and $i,j=1,2,3,4$ indices describe fundamental representations of $SO(6)_R$ and $SU(4)_R$ R-symmetry.
\\
\indent The component fields in a vector multiplet are given by a vector field $A_\mu$, four gaugini $\lambda^i$ and six scalars $\phi^m$. In general, the gravity multiplet can couple to an arbitrary number $n$ of vector multiplets. These vector
multiplets will be labeled by indices $a,b=1,\ldots, n$, so all fields in the vector multiplets will carry an additional index in the form 
\begin{equation}
(A^a_\mu,\lambda^{ia},\phi^{ma}).
\end{equation}  
Similar to the dilaton and the axion in the gravity multiplet, the $6n$ scalar fields $\phi^{ma}$ can be parametrized by $SO(6,n)/SO(6)\times SO(n)$ coset.
\\
\indent Fermionic fields and supersymmetry parameters transforming in fundamental representation of $SU(4)_R\sim SO(6)_R$ are subject to the chirality projections
\begin{equation}
\gamma_5\psi^i_\mu=\psi^i_\mu,\qquad \gamma_5\chi^i=-\chi^i,\qquad \gamma_5\lambda^i=\lambda^i
\end{equation}
while those transforming in anti-fundamental representation of $SU(4)_R$ satisfy
\begin{equation}
\gamma_5\psi_{\mu i}=-\psi_{\mu i},\qquad \gamma_5\chi_i=\chi_i,\qquad \gamma_5\lambda_i=-\lambda_i\, .
\end{equation}
\indent In general, all possible gaugings of a supergravity theory can be described by the embedding tensor. For the matter-coupled $N=4$ supergravity, supersymmetry requires that the embedding tensor can have only two non-vanishing components denoted by $\xi^{\alpha M}$ and $f_{\alpha MNP}$. Indices $\alpha=(+,-)$ and $M,N=(m,a)=1,\ldots, n+6$ describe fundamental representations of the global $SL(2,\mathbb{R})$ and $SO(6,n)$ symmetries, respectively. The electric vector fields $A^{+M}=(A^m_\mu,A^a_\mu)$, appearing in the ungauged Lagrangian with the usual Yang-Mills kinetic term, together with their magnetic dual $A^{-M}$ form a doublet under $SL(2,\mathbb{R})$ denoted by $A^{\alpha M}$. In addition, the embedding tensor needs to satisfy the quadratic constraint in order to define a consistent gauging. This ensures that the resulting gauge generators form a closed subalgebra of $SL(2,\mathbb{R})\times SO(6,n)$. We will not discuss all these details here but refer to the original construction in \cite{N4_gauged_SUGRA}, see also \cite{Eric_N4_4D,de_Roo_N4_4D,N4_Wagemans} for an earlier construction.
\\
\indent Consistent gauge groups can generally be embedded in both $SL(2,\mathbb{R})$ and
$SO(6,n)$ factors, and the magnetic vector fields can also participate in the gauging. However, each magnetic vector field must be accompanied by an auxiliary two-form field in order to remove the extra degrees of freedom. The analysis of the existence of maximally supersymmetric $AdS_4$ vacua given in \cite{AdS4_N4_Jan}, see also \cite{de_Roo_N4_4D,N4_Wagemans} for an earlier result, requires the gaugings to involve both electric and magnetic vector fields with the corresponding gauge groups embedded solely in $SO(6,n)$. This implies that both electric and magnetic components of $f_{\alpha MNP}$ must be non-vanishing and $\xi^{\alpha M}=0$. We also note that the magnetic components $f_{-MNP}$ are related to the duality phases first introduced in \cite{de_Roo_N4_4D,N4_Wagemans}. 
\\
\indent In this paper, we are interested in supersymmetric solutions that are asymptotic to $AdS_4$ vacua and involve only the metric and scalar fields. Accordingly, from now on, we will set all the other fields including $\xi^{\alpha M}$ to zero. The scalar coset $SL(2,\mathbb{R})/SO(2)\times SO(6,n)/SO(6)\times SO(n)$ can be described by the coset representative $\mc{V}_\alpha$ and $\mc{V}_M^{\phantom{M}A}$, respectively. With the definition 
\begin{equation}
\tau=\chi+ie^\phi,
\end{equation}
we will choose the explicit form of $\mc{V}_\alpha$ to be
\begin{equation}
\mc{V}_\alpha=e^{\frac{\phi}{2}}\left(
                                         \begin{array}{c}
                                           \chi+ie^{\phi} \\
                                           1 \\
                                         \end{array}
                                       \right).
\end{equation}
The coset representative $\mc{V}_M^{\phantom{M}A}$ transforms under the global $SO(6,n)$ and local $SO(6)\times SO(n)$ by left and right multiplications, respectively. This implies the splitting of the index $A=(m,a)$. We can then write the $SO(6,n)/SO(6)\times SO(n)$ coset representative as
\begin{equation}
\mc{V}_M^{\phantom{M}A}=(\mc{V}_M^{\phantom{M}m},\mc{V}_M^{\phantom{M}a}).
\end{equation}
Furthermore, as an element of $SO(6,n)$, the matrix $\mc{V}_M^{\phantom{M}A}$ also satisfies the relation
\begin{equation}
\eta_{MN}=-\mc{V}_M^{\phantom{M}m}\mc{V}_N^{\phantom{M}m}+\mc{V}_M^{\phantom{M}a}\mc{V}_N^{\phantom{M}a}
\end{equation}
with $\eta_{MN}=\textrm{diag}(-1,-1,-1,-1,-1,-1,1,\ldots,1)$ being the $SO(6,n)$ invariant tensor.
\\
\indent With $\xi^{\alpha M}=0$ and only the metric and scalars non-vanishing, the bosonic Lagrangian can simply be written as
\begin{equation}
e^{-1}\mc{L}=\frac{1}{2}R+\frac{1}{16}\pd_\mu M_{MN}\pd^\mu
M^{MN}-\frac{1}{4(\textrm{Im}\tau)^2}\pd_\mu \tau \pd^\mu \tau^*-V
\end{equation}
where $e=\sqrt{-g}$ is the vielbein determinant. The scalar potential is given in terms of the scalar coset representative and the embedding tensor by
\begin{eqnarray}
V&=&\frac{1}{16}\left[f_{\alpha MNP}f_{\beta
QRS}M^{\alpha\beta}\left[\frac{1}{3}M^{MQ}M^{NR}M^{PS}+\left(\frac{2}{3}\eta^{MQ}
-M^{MQ}\right)\eta^{NR}\eta^{PS}\right]\right.\nonumber \\
& &\left.-\frac{4}{9}f_{\alpha MNP}f_{\beta
QRS}\epsilon^{\alpha\beta}M^{MNPQRS}\right].
\end{eqnarray}
At this point, we should note that the components $f_{\alpha MNP}$ of the embedding tensor include the gauge coupling constants.
\\
\indent The $SO(6)\times SO(n)$ invariant and symmetric matrix $M_{MN}$ is defined by 
\begin{equation}
M_{MN}=\mc{V}_M^{\phantom{M}m}\mc{V}_N^{\phantom{M}m}+\mc{V}_M^{\phantom{M}a}\mc{V}_N^{\phantom{M}a}
\end{equation}
with its inverse denoted by $M^{MN}$. The tensor $M^{MNPQRS}$ is obtained from raising indices of $M_{MNPQRS}$ given by
\begin{equation}
M_{MNPQRS}=\epsilon_{mnpqrs}\mc{V}_{M}^{\phantom{M}m}\mc{V}_{N}^{\phantom{M}n}
\mc{V}_{P}^{\phantom{M}p}\mc{V}_{Q}^{\phantom{M}q}\mc{V}_{R}^{\phantom{M}r}\mc{V}_{S}^{\phantom{M}s}\label{M_6}\, .
\end{equation}
Similarly, $M^{\alpha\beta}$ is the inverse of the symmetric $2\times 2$ matrix $M_{\alpha\beta}$ defined by
\begin{equation}
M_{\alpha\beta}=\textrm{Re}(\mc{V}_\alpha\mc{V}^*_\beta).
\end{equation}
\indent Fermionic supersymmetry transformations are given by
\begin{eqnarray}
\delta\psi^i_\mu &=&2D_\mu \epsilon^i-\frac{2}{3}gA^{ij}_1\gamma_\mu
\epsilon_j,\\
\delta \chi^i &=&-\epsilon^{\alpha\beta}\mc{V}_\alpha D_\mu
\mc{V}_\beta\gamma^\mu \epsilon^i-\frac{4}{3}igA_2^{ij}\epsilon_j,\\
\delta \lambda^i_a&=&2i\mc{V}_a^{\phantom{a}M}D_\mu
\mc{V}_M^{\phantom{M}ij}\gamma^\mu\epsilon_j-2igA_{2aj}^{\phantom{2aj}i}\epsilon^j\,
.
\end{eqnarray}
The fermion shift matrices are in turn defined by
\begin{eqnarray}
A_1^{ij}&=&\epsilon^{\alpha\beta}(\mc{V}_\alpha)^*\mc{V}_{kl}^{\phantom{kl}M}\mc{V}_N^{\phantom{N}ik}
\mc{V}_P^{\phantom{P}jl}f_{\beta M}^{\phantom{\beta M}NP},\nonumber
\\
A_2^{ij}&=&\epsilon^{\alpha\beta}\mc{V}_\alpha\mc{V}_{kl}^{\phantom{kl}M}\mc{V}_N^{\phantom{N}ik}
\mc{V}_P^{\phantom{P}jl}f_{\beta M}^{\phantom{\beta M}NP},\nonumber
\\
A_{2ai}^{\phantom{2ai}j}&=&\epsilon^{\alpha\beta}\mc{V}_\alpha
{\mc{V}_a}^M{\mc{V}_{ik}}^N\mc{V}_P^{\phantom{P}jk}f_{\beta
MN}^{\phantom{\beta MN}P}
\end{eqnarray}
where $\mc{V}_M^{\phantom{M}ij}$ is defined in terms of the 't Hooft
symbols $G^{ij}_m$ and $\mc{V}_M^{\phantom{M}m}$ as
\begin{equation}
\mc{V}_M^{\phantom{M}ij}=\frac{1}{2}\mc{V}_M^{\phantom{M}m}G^{ij}_m\, .
\end{equation}
Similarly, the inverse elements ${\mc{V}_{ij}}^M$ can be written as
\begin{equation}
{\mc{V}_{ij}}^M=-\frac{1}{2}{\mc{V}_{m}}^M(G^{ij}_m)^*\,
.
\end{equation}
$G^{ij}_m$ satisfy the relations
\begin{equation}
G_{mij}=(G^{ij}_m)^*=\frac{1}{2}\epsilon_{ijkl}G^{kl}_m\, .
\end{equation}
An explicit representation of these matrices can be chosen as
\begin{eqnarray}
G_1^{ij}&=&\left[
                     \begin{array}{cccc}
                       0 & 1 & 0 & 0 \\
                       -1 & 0 & 0 & 0 \\
                       0 & 0 & 0  & 1\\
                        0 & 0 & -1  & 0\\
                     \end{array}
                   \right],\qquad
G_2^{ij}=\left[
                     \begin{array}{cccc}
                       0 & 0 & 1 & 0\\
                       0 & 0 & 0 & -1\\
                       -1 & 0 & 0  & 0\\
                        0 & 1 & 0  & 0\\
                     \end{array}
                   \right],\nonumber \\
G_3^{ij}&=&\left[
                     \begin{array}{cccc}
                       0 & 0 & 0 & 1\\
                       0 & 0 & 1 & 0\\
                       0 & -1 & 0  & 0\\
                        -1 & 0 & 0  & 0\\
                     \end{array}
                   \right],\qquad
G_4^{ij}=\left[
                     \begin{array}{cccc}
                       0 & i & 0 & 0\\
                       -i & 0 & 0 & 0\\
                       0 & 0 & 0  & -i\\
                        0 & 0 & i  & 0\\
                     \end{array}
                   \right],\nonumber \\
G_5^{ij}&=&\left[
                     \begin{array}{cccc}
                       0 & 0 & i & 0\\
                       0 & 0 & 0 & i\\
                       -i & 0 & 0  & 0\\
                        0 & -i & 0  & 0\\
                     \end{array}
                   \right],\qquad
G_6^{ij}=\left[
                     \begin{array}{cccc}
                       0 & 0 & 0 & i\\
                       0 & 0 & -i & 0\\
                       0 & i & 0  & 0\\
                        -i & 0 & 0  & 0\\
                     \end{array}
                   \right].  
\end{eqnarray}
\indent Finally, we also note that the scalar potential can be written in terms of the fermion shift matrices $A_1$ and $A_2$ as
\begin{equation}
V=-\frac{1}{3}A^{ij}_1A_{1ij}+\frac{1}{9}A^{ij}_2A_{2ij}+\frac{1}{2}A_{2ai}^{\phantom{2ai}j}
A_{2a\phantom{i}j}^{\phantom{2a}i}\, .
\end{equation}
It is useful to remark here that upper and lower $i,j,\ldots$ indices are related by complex conjugation.

\subsection{$SO(4)\times SO(4)$ gauge group and $SO(2)\times SO(2)\times SO(2)\times SO(2)$ truncation}
We now consider the case of $n=6$ vector multiplets with $SO(4)\times SO(4)$ gauge group as constructed in \cite{dS_Roest}. With a slightly different notation for various gauge coupling constants, the corresponding embedding tensor has the following non-vanishing components 
\begin{eqnarray}
f_{+\hat{m}\hat{n}\hat{p}}&=&g_1\epsilon_{\hat{m}\hat{n}\hat{p}},\qquad f_{+\hat{a}\hat{b}\hat{c}}=\tilde{g}_1\epsilon_{\hat{a}\hat{b}\hat{c}},\nonumber \\
f_{-\tilde{m}\tilde{n}\tilde{p}}&=&g_2\epsilon_{\tilde{m}\tilde{n}\tilde{p}},\qquad f_{-\tilde{a}\tilde{b}\tilde{c}}=\tilde{g}_2\epsilon_{\tilde{a}\tilde{b}\tilde{c}},\label{embedding_tensor}
\end{eqnarray}
in which we have used the convention on the $SO(6,6)$ fundamental index as $M=(m,a)=(\hat{m},\tilde{m},\hat{a},\tilde{a})$ with $\hat{m}=1,2,3$, $\tilde{m}=4,5,6$, $\hat{a}=7,8,9$ and $\tilde{a}=10,11,12$. We also note that the two $SO(4)$ factors are electrically and magnetically embedded in $SO(6,6)$ and will be denoted by $SO(4)_+\times SO(4)_-$. In terms of the $SO(3)$ factors given by the embedding tensor in \eqref{embedding_tensor}, we will write the gauge group as $SO(3)_+\times SO(3)_-\times SO(3)_+\times SO(3)_-$ with the first two factors embedded in the $SU(4)_R\sim SO(6)_R$.  
\\
\indent To parametrize the $SO(6,6)/SO(6)\times SO(6)$ coset representative, we first define $SO(6,n)$ generators in the fundamental representation by
\begin{equation}
(t_{MN})_P^{\phantom{P}Q}=2\delta^Q_{[M}\eta_{N]P}\, .
\end{equation}
We then identify the $SO(6,6)$ non-compact generators as
\begin{equation}
Y_{ma}=t_{m,a+6}\, .
\end{equation}
The truncation of this gauged supergravity to $SO(4)_{\textrm{diag}}\sim SO(3)_{\textrm{diag}}\times SO(3)_{\textrm{diag}}$ singlet scalars has already been studied in \cite{4D_N4_flows} in which a number of supersymmetric $AdS_4$ critical points and domain walls interpolating between them have been given. 
\\
\indent In the present case, we are interested in another truncation to $SO(2)\times SO(2)\times SO(2)\times SO(2)$ invariant scalars. For later convenience, we also note that the gauge generators are given by 
\begin{equation}
X_{\alpha M}={f_{\alpha M}}^{NP}t_{NP}\, .
\end{equation}
Each $SO(2)$ factor is embedded in the $SO(3)$ factor of $SO(3)_+\times SO(3)_-\times SO(3)_+\times SO(3)_-\sim SO(4)_+\times SO(4)_-$ as $\mathbf{3}\rightarrow \mathbf{2}+\mathbf{1}$. To identify these singlets, we first consider the transformation of the scalars under the $SO(2)\times SO(2)\times SO(2)\times SO(2)$ subgroup. The $36$ scalars transform under the compact subgroup $SO(6)\times SO(6)$ as $(\mathbf{6},\mathbf{6})$. The fundamental representation $\mathbf{6}$ of $SO(6)$ is in turn decomposed as $(\mathbf{3},\mathbf{1})+(\mathbf{1},\mathbf{3})$ under $SO(3)\times SO(3)$. With the aforementioned embedding of $SO(2)$ in $SO(3)$, we find the transformation of all scalars under $SO(2)\times SO(2)\times SO(2)\times SO(2)$
\begin{equation}
(\mathbf{6},\mathbf{6})\rightarrow (\mathbf{2}+\mathbf{1},\mathbf{2}+\mathbf{1},\mathbf{1},\mathbf{1})+(\mathbf{2}+\mathbf{1},\mathbf{1},\mathbf{1},\mathbf{2}+\mathbf{1})+(\mathbf{1},\mathbf{2}+\mathbf{1},\mathbf{2}+\mathbf{1},\mathbf{1})+(\mathbf{1},\mathbf{1},\mathbf{2}+\mathbf{1},\mathbf{2}+\mathbf{1})
\end{equation}
from which we can easily see that there are four singlets corresponding to the representation $(\mathbf{1},\mathbf{1},\mathbf{1},\mathbf{1})$. With the gauge generators obtained from the embedding tensor given in \eqref{embedding_tensor}, we can choose the $SO(2)\times SO(2)\times SO(2)\times SO(2)$ generators to be $X_{+3}$, $X_{-6}$, $X_{+9}$ and $X_{-12}$. We then identify non-compact generators corresponding to these singlets as $Y_{33}$, $Y_{36}$, $Y_{63}$ and $Y_{66}$ in term of which the coset representative can be written as
\begin{equation}
\mc{V}=e^{\phi_1 Y_{33}}e^{\phi_2 Y_{36}}e^{\phi_3 Y_{63}}e^{\phi_4 Y_{66}}\, .\label{SO2_4_coset}
\end{equation}
Together with the dilaton and axion from the gravity multiplet which are $SO(4)\times SO(4)$ singlets, there are six scalars in the $SO(2)\times SO(2)\times SO(2)\times SO(2)$ sector.
\\
\indent By a straightforward computation, we find a very simple scalar potential
\begin{equation}
V=-\frac{1}{2}e^{-\phi}(g_1^2+e^{2\phi}g_2^2+g_2^2\chi^2)-2g_1g_2\cosh\phi_1\cosh\phi_2\cosh\phi_3\cosh\phi_4\, .\label{Potential}
\end{equation}
It can be readily verified that this potential admits a unique $AdS_4$ critical point at
\begin{equation}
\phi=\ln\left[\frac{g_1}{g_2}\right]\qquad \textrm{and}\qquad \phi_1=\phi_2=\phi_3=\phi_4=\chi=0\, .
\end{equation}
The cosmological constant and $AdS_4$ radius are given by
\begin{equation}
V_0=-3g_1g_2\qquad \textrm{and}\qquad L=\sqrt{-\frac{3}{V_0}}=\frac{1}{\sqrt{g_1g_2}}\, .
\end{equation}
This is the maximally $N=4$ supersymmetric $AdS_4$ vacuum preserving the full $SO(4)\times SO(4)$ gauge symmetry. By shifting the dilaton $\phi$ by a constant, we can choose $g_2=g_1=g$ which make the dilaton vanish at the $AdS_4$ critical point. It is useful to note that all scalars have mass $m^2L^2=-2$ at this vacuum and are dual to operators of dimensions $\Delta=1,2$ in the dual $N=4$ SCFT, see more detail in \cite{4D_N4_flows}. 

\subsection{Superpotential and BPS equations}
We are interested in two types of supersymmetric solutions in the forms of holographic RG flows and Janus configurations. The former is described by the standard (flat) domain wall with the metric
\begin{equation}
ds^2=e^{2A(r)}dx^2_{1,2}+dr^2\label{DW_metric}
\end{equation}
while the latter takes the form of a curved domain wall with an $AdS_3$-sliced world-volume corresponding to the metric
\begin{equation}
ds^2=e^{2A(r)}(e^{\frac{2\xi}{\ell}}dx^2_{1,1}+d\xi^2)+dr^2\, .\label{Janus_metric}
\end{equation}
It should be noted that the Janus ansatz \eqref{Janus_metric} is obtained from the domain wall metric \eqref{DW_metric} by replacing the three-dimensional Minkowski metric $dx^2_{1,2}$ by the metric on $AdS_3$ with radius $\ell$. The flat domain wall is in turn recovered from the Janus metric \eqref{Janus_metric} in the limit $\ell \rightarrow \infty$.
\\
\indent To preserve the isometry of $dx_{1,2}^2$ or the $AdS_3$ metric, scalar fields can depend only on the radial coordinate $r$. In the $SO(2)\times SO(2)\times SO(2)\times SO(2)$ truncation, the kinetic term for all the six scalars $\Phi^r=(\phi,\chi,\phi_1,\phi_2,\phi_3,\phi_4)$, with $r,s=1,2,\ldots, 6$, takes the form
\begin{eqnarray}
\mc{L}_{\textrm{kin}}&=&\frac{1}{2}G_{rs}{\Phi^r}'{\Phi^s}'\nonumber \\
&=&-\frac{1}{4}(\phi'^2+e^{-2\phi}\chi'^2)-\frac{1}{16}\left[6+\cosh2(\phi_2-\phi_3)\right. \nonumber \\
& &
\left.+\cosh2(\phi_2+\phi_3)+2\cosh2\phi_4(\cosh2\phi_2\cosh2\phi_3-1)\right]\phi'^2_1\nonumber \\
& &-\cosh\phi_2\cosh\phi_4\sinh\phi_3\sinh\phi_4\phi'_1\phi'_2-\cosh\phi_3\cosh\phi_4\sinh\phi_2\sinh\phi_4\phi'_1\phi'_3\nonumber \\
& &+\sinh\phi_2\sinh\phi_3\phi'_1\phi'_4-\frac{1}{2}\cosh^2\phi_4\phi'^2_2-\frac{1}{2}\cosh^2\phi_4\phi'^2_3-\frac{1}{2}\phi'^2_4
 \end{eqnarray} 
in which we have introduced a symmetric matrix $G_{rs}$. Throughout the paper, we will denote the $r$-derivative by $'$.
\\
\indent We now analyze the BPS equations obtained by setting fermionic supersymmetry transformations to zero. We use Majorana representation for space-time gamma matrices with all $\gamma^{\hat{\mu}}$ real and $\gamma_5$ purely imaginary. This implies that left and right chiralities of fermions are simply related to each other by complex conjugation. To proceed, we first note that for both types of solutions, the variations $\delta \chi^i$ and $\delta \lambda^i_a$ contain $\gamma^{\hat{r}}$ matrix due to the $r$-dependence of scalar fields. We then impose the following projector
\begin{equation}
\gamma^{\hat{r}}\epsilon^i=e^{i\Lambda}\epsilon_i\label{gamma_r_pro}
\end{equation}
with an $r$-dependent phase $\Lambda$. This projector relates the two chiralities of $\epsilon^i$ and breaks half of the supersymmetry.
\\
\indent To complete the analysis, we consider the variation of the gravitini $\delta \psi^i_\mu$. With the coset representative \eqref{SO2_4_coset}, we obtain the $A_1^{ij}$ tensor of the form
\begin{equation}
A^{ij}_1=\textrm{diag}(\alpha_+,\alpha_-,\alpha_-,\alpha_+)
\end{equation}
with
\begin{eqnarray}
\alpha_\pm&=&\frac{3}{4}e^{-\frac{\phi}{2}}\left[(g_2e^\phi\cosh\phi_3\pm ig_1\sinh\phi_1\sinh\phi_3)\cosh\phi_4 \right. \nonumber \\
& &+g_1\cosh\phi_1(\cosh\phi_2\pm i\sinh\phi_2\sinh\phi_4)+\nonumber \\
& &\left. +ig_2\chi\cosh\phi_3\cosh\phi_4\right].
\end{eqnarray}
For $\chi=0$, $\alpha_+$ and $\alpha_-$ are complex conjugate of each other. The eigenvalue of $A_1^{ij}$ tensor corresponding to unbroken supersymmetry will give a superpotential in term of which the scalar potential can be written. From the structure of $A_1^{ij}$ given above, we find that in general, solutions within the $SO(2)\times SO(2)\times SO(2)\times SO(2)$ truncation will preserve at most $N=2$ supersymmetry corresponding to $\epsilon^{1,4}$ or $\epsilon^{2,3}$ depending on which eigenvalue among $\alpha_+$ and $\alpha_-$ is chosen to be the superpotential. 
\\
\indent By defining the superpotential $\mc{W}$ as
\begin{equation}
\mc{W}=\frac{2}{3}\alpha
  \end{equation}  
and taking into account the projector \eqref{gamma_r_pro}, we can write the variations $\delta \psi^i_\mu$ for $\mu=0,1,2$ in the RG flow case as
\begin{equation}
A'e^{i\Lambda}-\mc{W}=0\, .\label{RG_eq1}
\end{equation}
This condition is satisfied along $\epsilon^{1,4}$ and $\epsilon^{2,3}$ for $\mc{W}=\mc{W}_\pm=\frac{2}{3}\alpha_\pm$, respectively. In the following analysis, we will consider only $\mc{W}=\mc{W}_+$. The final results show that the other choice $\mc{W}=\mc{W}_-$ is related to this by a sign change in $\chi$, so the two options are equivalent. Equation \eqref{RG_eq1} implies that
\begin{equation}
A'=\pm W=\pm|\mc{W}| \qquad \textrm{and}\qquad e^{i\Lambda}=\pm\frac{\mc{W}}{W}\, .\label{RG_flow_phase}
\end{equation}
Using the explicit phase factor $e^{i\Lambda}$ in term of $\mc{W}$ in $\delta \chi^i$ and $\delta \lambda^i_a$ equations, we obtain the BPS equations for scalar fields. In what follows, we will choose the upper sign choice in order to relate the $AdS_4$ vacuum to the limit $r\rightarrow \infty$.
\\
\indent Finally, the variation $\delta\psi_r^i$ leads to 
\begin{equation}
2{\epsilon^{\hat{i}}}'-\mc{W}\gamma_{\hat{r}}\epsilon_{\hat{i}}=0,\qquad \hat{i}=1,4
\end{equation}
which, by the projector \eqref{gamma_r_pro} and equation \eqref{RG_eq1}, gives the Killing spinors of the form
\begin{equation}
\epsilon^{\hat{i}}=e^{\frac{A}{2}}\epsilon^{\hat{i}}_0
  \end{equation}  
for constant spinors $\epsilon_0^{\hat{i}}$ satisfying \eqref{gamma_r_pro}.   
\\
\indent For Janus solutions, we can perform a similar analysis with some modifications. The variations $\delta\psi^{\hat{i}}_{0,1}$ and $\delta \psi^{\hat{i}}_\xi$ lead to the following conditions
\begin{eqnarray}
\left(A'\gamma^{\hat{r}}+\frac{e^{-A}}{\ell}\gamma^{\hat{\xi}}\right)\epsilon^{\hat{i}}-\mc{W}\epsilon_{\hat{i}}&=&0,\label{Janus_eq1}\\
2\pd_{\xi}\epsilon^{\hat{i}}+A'e^A\gamma^{\hat{\xi}}\gamma^{\hat{r}}\epsilon^{\hat{i}}-e^A\mc{W}\gamma^{\hat{\xi}}\epsilon_{\hat{i}}&=&0\, .\label{Janus_eq2}
\end{eqnarray}
The analysis closely follows that given in \cite{warner_Janus} to which we refer for more detail. 
\\
\indent Equation \eqref{Janus_eq1} implies the integrability condition
\begin{equation}
A'^2+\frac{e^{-2A}}{\ell^2}=W^2\, .
\end{equation}
Using equation \eqref{Janus_eq1} in \eqref{Janus_eq2}, we find
\begin{equation}
2\pd_\xi \epsilon^{\hat{i}}=\frac{1}{\ell}\epsilon^{\hat{i}}
\end{equation}
which gives
\begin{equation}
\epsilon^{\hat{i}}=e^{\frac{\xi}{2\ell}}\tilde{\epsilon}^{\hat{i}}
\end{equation}
with $\tilde{\epsilon}^{\hat{i}}$ being $\xi$-independent spinors.
\\
\indent It should be noted that condition \eqref{Janus_eq1} takes the form of a projector on $\epsilon^{\hat{i}}$ which must be compatible with the projector \eqref{gamma_r_pro} used in the scalar flow equations. Following \cite{warner_Janus}, we impose the projector 
\begin{equation}
\gamma^{\hat{\xi}}\epsilon^{\hat{i}}=i\kappa e^{i\Lambda}\epsilon_{\hat{i}}\label{gamma_xi_pro}
\end{equation}
with $\kappa^2=1$. Using equations \eqref{gamma_r_pro} and \eqref{gamma_xi_pro} in \eqref{Janus_eq1}, we arrive at
\begin{equation} 
\left(A'+\frac{i\kappa}{\ell}e^{-A}\right)e^{i\Lambda}=\mc{W}
\end{equation}
which can be used to determine the phase $e^{i\Lambda}$ in terms of $\mc{W}$ and $A$. We note again that in the limit $\ell\rightarrow \infty$, the phase factor reduces to that given in \eqref{RG_flow_phase}.
\\
\indent Finally, with all the previous results, the condition $\delta\psi^{\hat{i}}_r=0$ gives
\begin{equation}
2\pd_r\epsilon^{\hat{i}}=A'\epsilon^{\hat{i}}+\frac{i\kappa}{\ell}e^{-A}\epsilon^{\hat{i}}\, .
\end{equation}
By redefining the Killing spinors with the phase factor $e^{\frac{i\Lambda}{2}}$, this condition implies the following form of the Killing spinors
\begin{equation}
\epsilon^{\hat{i}}=e^{\frac{A}{2}+\frac{r}{2\ell}+\frac{i\Lambda}{2}}\epsilon_0^{\hat{i}}
\end{equation}
in which $\epsilon_0^{\hat{i}}$ can include an $r$-dependent phase and satisfy the $\gamma^{\hat{r}}$ and $\gamma^{\hat{\xi}}$ projectors without the phase $e^{i\Lambda}$
\begin{equation}
 \gamma^{\hat{r}}\epsilon^{\hat{i}}_0=\epsilon_{0\hat{i}}\qquad \textrm{and}\qquad \gamma^{\hat{\xi}}\epsilon^{\hat{i}}_0=i\kappa\epsilon_{0\hat{i}}\, .
\end{equation}   
We also note that the constant $\kappa=\pm1$ is related to the chirality of the Killing spinors on the two-dimensional interface dual to the $AdS_3$ slice.
\\
\indent With all these results, we will look for supersymmetric RG flow and Janus solutions within the $SO(2)\times SO(2)\times SO(2)\times SO(2)$ truncation of the $SO(4)\times SO(4)$ $N=4$ gauged supergravity.

\section{Holographic RG flows}\label{RG_flow}
We are now in a position to present an explicit form of the BPS equations given in the previous section and look for possible solutions. In this section, we consider RG flow solutions. Since there is only one supersymmetric $AdS_4$ critical point dual to a single conformal fixed point, the solutions will describe RG flows from the dual $N=4$ SCFT to non-conformal phases in the IR. This type of solutions in the case of pure $N=4$ gauged supergravity, with all scalars but the dilaton and axion vanishing, has already been studied in \cite{4D_N4_flows}. In this paper, we will find solutions in the matter-coupled $N=4$ gauged supergravity. 
\\
\indent As previously mentioned, the flow equations for scalars can be obtained by using the projector \eqref{gamma_r_pro} and the phase $e^{i\Lambda}$ given in \eqref{RG_flow_phase}. Within the $SO(2)\times SO(2)\times SO(2)\times SO(2)$ truncation, the variation $\delta\chi^i$ reduces to a single complex equation which can be straightforwardly solved for $\phi'$ and $\chi'$. However, it turns out that consistency of the equations from the variation $\delta \lambda^i_a$ requires some scalars to vanish. This is similar to the $SO(3)\times SO(3)$ truncation studied in \cite{4D_N4_flows} in which the BPS equations from $\delta \lambda^i_a$ imply $\chi=0$, and non-vanishing $\chi$ is possible only when all scalars from the vector multiplets are set to zero. In that case, the axion and vector multiplet scalars cannot be turned on simultaneously.
\\
\indent In the present case, the constraint is somewhat weaker. We find that a consistent set of BPS equations that are compatible with the second-order field equations can be obtained only for two of the scalars $\phi_i$, $i=1,2,3,4$ vanishing. This leads to $N=2$ supersymmetric solutions. In addition, if the axion $\chi$ vanishes, supersymmetry is enhanced to $N=4$. As expected, truncating out all scalars from the vector multiplets also leads to $N=4$ supersymmetry. We then see that in the presence of both axion and vector multiplet scalars, the solutions break some supersymmetry. We also point out that this is not possible in the $SO(3)\times SO(3)$ truncation considered in \cite{4D_N4_flows} since in that case the $A^{ij}_1$ tensor is proportional to the identity matrix. Therefore, partial supersymmetry breaking is not possible there. 

\subsection{$N=4$ holographic RG flow with $SO(2)\times SO(3)\times SO(2)\times SO(2)$ symmetry}
We first consider $N=4$ solutions with $\chi=0$ and $SO(2)\times SO(3)\times SO(2)\times SO(2)$ symmetry. In this case, the $SO(4)$ subgroup of the $SO(6)$ R-symmetry is broken to $SO(2)\times SO(3)$ while the remaining $SO(2)\times SO(2)$ factor is a subgroup of $SO(3)\times SO(3)\sim SO(4)$ symmetry of the vector multiplets. There are two possibilities namely setting $\phi_1=\phi_2=0$ or $\phi_3=\phi_4=0$. The two choices are related to each other by interchanging $e^{\phi}$ with $e^{-\phi}$ factors. Therefore, we will consider only the solution with $\phi_3=\phi_4=0$.
\\
\indent The superpotential is real and given by
\begin{equation}
W=W_+=W_-=\frac{1}{2}(g_2e^{\frac{\phi}{2}}+g_1e^{-\frac{\phi}{2}}\cosh\phi_1\cosh\phi_2).
\end{equation}
The resulting BPS equations read
\begin{eqnarray}
\phi'&=&-4\frac{\pd W}{\pd \phi}=g_1e^{-\frac{\phi}{2}}\cosh\phi_1\cosh\phi_2-g_2e^{\frac{\phi}{2}},\\
\phi_1'&=&-2\textrm{sech}^2\phi_2\frac{\pd W}{\pd \phi_1}=-g_1e^{-\frac{\phi}{2}}\textrm{sech}\phi_2\sinh\phi_1,\\
\phi_2'&=&-2\frac{\pd W}{\pd \phi_2}=-g_1e^{-\frac{\phi}{2}}\cosh\phi_1\sinh\phi_2,\\
A'&=&W=\frac{1}{2}(g_2e^{\frac{\phi}{2}}+g_1e^{-\frac{\phi}{2}}\cosh\phi_1\cosh\phi_2).
\end{eqnarray}
To find the solution, we first take the following combination
\begin{equation}
\frac{d\phi_1}{d\phi_2}=\textrm{csch}\phi_2\textrm{sech}\phi_2\tanh\phi_1
\end{equation}
with the solution
\begin{equation}
\sinh\phi_1=\tan\varphi_0\tanh\phi_2
\end{equation}
for a constant $\varphi_0$. Similarly, taking the combination between $\phi'$ and $\phi_2'$ equations together with this result leads to
\begin{equation}
\frac{d\phi}{d\phi_2}=\frac{g_2e^\phi\textrm{csch}\phi_2}{g_1\sqrt{1+\tan\varphi_0^2\tanh^2\phi_2}}-\coth\phi_2\, .
\end{equation}
The solution to this equation is given by
\begin{equation}
\phi=\ln\left[\frac{2\sqrt{2}g_1\cos^2\varphi_0\textrm{csch}\phi_2(\cos\varphi_0\textrm{csch}\phi_2\sqrt{\cos2\varphi_0
+\cosh2\phi_2}-C)}{g_2[1-2C^2+\cos2\varphi_0+2\cos^2\varphi_0(\coth^2\phi_2+\cos2\varphi_0\textrm{csch}^2\phi_2)]}\right].
\end{equation}
Using these results and redefining the radial coordinate to $\rho$ given by $\frac{d\rho}{dr}=e^{-\frac{\phi}{2}}$, we obtain the solution for $\phi_2(\rho)$ of the form
\begin{equation}
\cosh2\phi_2=2\cos^2\varphi_0\tanh\left[g_1\sec\varphi_0(\rho-\rho_0)\right]-\cos2\varphi_0
\end{equation}
with an integration constant $\rho_0$.
\\
\indent Finally, taking a linear combination $2A'+\phi'$, we find
\begin{equation}
2\frac{dA}{d\phi_2}+\frac{d\phi}{d\phi_2}=-\frac{2(e^{2\phi_2}+1)}{e^{2\phi_2}-1}
\end{equation}
with the solution
\begin{equation}
A=\phi_2-\frac{1}{2}\phi-\ln(1-e^{2\phi_2}).
\end{equation}
We have neglected an additive integration constant in $A$ since this can be removed by rescaling coordinates on $dx^2_{1,2}$.
\\
\indent We now look at asymptotic behaviors of the solution. For convenience, we first set $g_2=g_1$ to bring the $AdS_4$ critical point to the origin of the scalar manifold. As $r\rightarrow \infty$, we find
\begin{equation}
\phi\sim \phi_1\sim \phi_2\sim e^{-g_1r}\sim e^{-\frac{r}{L}},\qquad A\sim g_1r\sim \frac{r}{L}\, .
 \end{equation} 
This gives the supersymmetric $AdS_4$ vacuum and indicates that all scalars are dual to relevant operators of dimensions $\Delta=1,2$. 
\\
\indent As usual for a flow to non-conformal phases, the solution is singular for a finite value of $\rho$. We find that, for $C<0$, the above solution is singular at $\rho=\rho_0+\rho_*$ with
\begin{equation}
\rho_*=\frac{1}{g_1}\coth^{-1}\left[\sec\varphi_0\sqrt{\frac{\cos2\varphi_0+\cosh\left(2\coth^{-1}\sqrt{1-\sec^2\varphi_0+\frac{1}{2}C^2\sec^4\varphi_0}\right)}{2}}\right].
\end{equation}
Near the singularity, we find
\begin{eqnarray}
& &\phi_2=\phi_*=\coth^{-1}\sqrt{1-\sec^2\varphi_0+\frac{1}{2}C^2\sec^4\varphi_0},\nonumber \\
& &\phi_1=\sinh^{-1}(\tan\varphi_0\tanh\phi_*),\qquad \phi\sim -\ln (\phi_2-\phi_*),\nonumber \\ 
& & A\sim -\frac{\phi}{2}\sim \frac{1}{2}\ln (\phi_2-\phi_*)
\end{eqnarray}
and
\begin{equation}
V\sim -g_2^2e^{\phi}\rightarrow -\infty\, .
\end{equation}
We see that near the singularity both $\phi_1$ and $\phi_2$ are constant, and the singularity is physical by the criterion given in \cite{Gubser_singularity}. Therefore, the solution describes an RG flow from $N=4$ SCFT to a non-conformal phase in the IR with the $N=4$ superconformal symmetry broken to $N=4$ Poincare supersymmetry in three dimensions. The $SO(4)\times SO(4)$ symmetry is also broken to $SO(2)\times SO(3)\times SO(2)\times SO(2)$ along the flows.  

\subsection{$N=4$ holographic RG flow with $SO(2)\times SO(2)\times SO(2)\times SO(2)$ symmetry}
Unlike the previous case, if we set $\phi_2=\phi_3=0$ or $\phi_1=\phi_4=0$ along with $\chi=0$, we obtain $N=4$ solutions with only $SO(2)\times SO(2)\times SO(2)\times SO(2)$ symmetry. The two choices give rise to equivalent results with scalars $(\phi_1,\phi_4)$ and $(\phi_2,\phi_3)$ interchanged. We will consider only the $\phi_2=\phi_3=0$ case for definiteness.
\\
\indent Setting $\chi=\phi_2=\phi_3=0$, we again find a real superpotential 
\begin{equation}
\mc{W}=\frac{1}{2}(g_1e^{-\frac{\phi}{2}}\cosh\phi_1+g_2e^{\frac{\phi}{2}}\cosh\phi_4).
\end{equation}  
By repeating the same procedure as in the previous case, we find the following BPS equations
\begin{eqnarray}
A'&=&W=\frac{1}{2}(g_1e^{-\frac{\phi}{2}}\cosh\phi_1+g_2e^{\frac{\phi}{2}}\cosh\phi_4),\\
\phi'&=&-\frac{\pd W}{\pd\phi}=g_1e^{-\frac{\phi}{2}}\cosh\phi_1-g_2e^{\frac{\phi}{2}}\cosh\phi_4,\\
\phi_1'&=&-2\frac{\pd W}{\pd\phi_1}=-g_1e^{-\frac{\phi}{2}}\sinh\phi_1,\\
\phi_4'&=&-2\frac{\pd W}{\pd\phi_4}=-g_2e^{\frac{\phi}{2}}\sinh\phi_4\, .
\end{eqnarray}
Unlike the previous case, solving these equations is slightly more complicated due to the different dilaton prefactors in $\phi_1'$ and $\phi_4'$ equations. However, with some effort, we can find an analytic solution to these equations.  
\\
\indent We first treat $\phi_4$ as an independent variable and take a combination 
\begin{equation}
\frac{d\phi_1}{d\phi_4}=\frac{g_1}{g_2}e^{-\phi}\textrm{csch}\phi_4\sinh\phi_1\, .
\end{equation}
By changing to a new variable $\varphi$ defined by
\begin{equation}
\phi_1=\ln \sinh\frac{\varphi}{2}-\ln\cosh\frac{\varphi}{2},
\end{equation}
we can rewrite the above equation as
\begin{equation}
\frac{d\varphi}{d\phi_4}=\frac{g_1}{g_2}e^{-\phi}\textrm{csch}\phi_4\, .\label{dvarphi_phi4}
\end{equation}
We then combine $\phi'$ and $\phi_4'$ equations with the $e^{-\phi}$ factor obtained from \eqref{dvarphi_phi4} and arrive at
\begin{equation}
\frac{d\phi}{d\phi_4}=\coth\phi_4+\coth\varphi\frac{d\varphi}{d\phi_4}
\end{equation}
with the solution
\begin{equation}
\phi=\phi_0+\ln \sinh\phi_4+\ln \sinh\varphi\, .
\end{equation}
We can now use this solution in \eqref{dvarphi_phi4} and find the solution
\begin{equation}
\cosh\varphi=\frac{g_1}{g_2}e^{-\phi_0}\coth\phi_4\, .
\end{equation}
Taking a linear combination $2A'-\phi'$ gives
\begin{equation}
2\frac{dA}{d\phi_4}-\frac{d\phi}{d\phi_4}=-2\coth\phi_4
\end{equation}
with the solution
\begin{equation}
A=\frac{1}{2}\phi-\ln \sinh\phi_4\, .
\end{equation}
\indent Finally, changing $r$ to a new coordinate $\rho$ given by $\frac{d\rho}{dr}=e^{\frac{\phi}{2}}$, we can readily find $\phi_4(\rho)$ solution
\begin{equation}
\phi_4=\ln(1+e^{-g_2(\rho-\rho_0)})-\ln(1-e^{-g_2(\rho-\rho_0)}).
\end{equation}
\indent As in the previous case, for $r\rightarrow\infty$, we find
\begin{equation}
\phi\sim \phi_1\sim\phi_4\sim e^{-g_1r}\sim e^{-\frac{r}{L}}
\end{equation}
in which we have taken $g_2=g_1$ for convenience. This is an asymptotic $AdS_4$ geometry.
\\
\indent As $\rho\rightarrow \rho_0$, the solution is singular with
\begin{eqnarray}
& &\phi_1\sim \textrm{constant},\qquad \phi_4\sim \ln (\rho-\rho_0),\nonumber\\
& &\phi\sim -\ln (\rho-\rho_0),\qquad A\sim \frac{1}{2}\ln(\rho-\rho_0).
\end{eqnarray}
Unlike the previous case, the scalars from vector multiplets are not all constant near the singularity. Using the scalar potential given in \eqref{Potential}, we find that, as $\rho\rightarrow \rho_0$, 
\begin{equation}
V\sim -\frac{g_1^2}{(\rho-\rho_0)}\rightarrow -\infty\, .
\end{equation}
The singularity is again physically acceptable, and the solution could be interpreted as an RG flow from the $N=4$ SCFT to a non-conformal phase in the IR driven by relevant operators of dimensions $\Delta=1,2$.

\subsection{$N=2$ holographic RG flow with $SO(2)\times SO(2)\times SO(3)\times SO(2)$ symmetry}
\begin{figure}
  \centering
  \begin{subfigure}[b]{0.45\linewidth}
    \includegraphics[width=\linewidth]{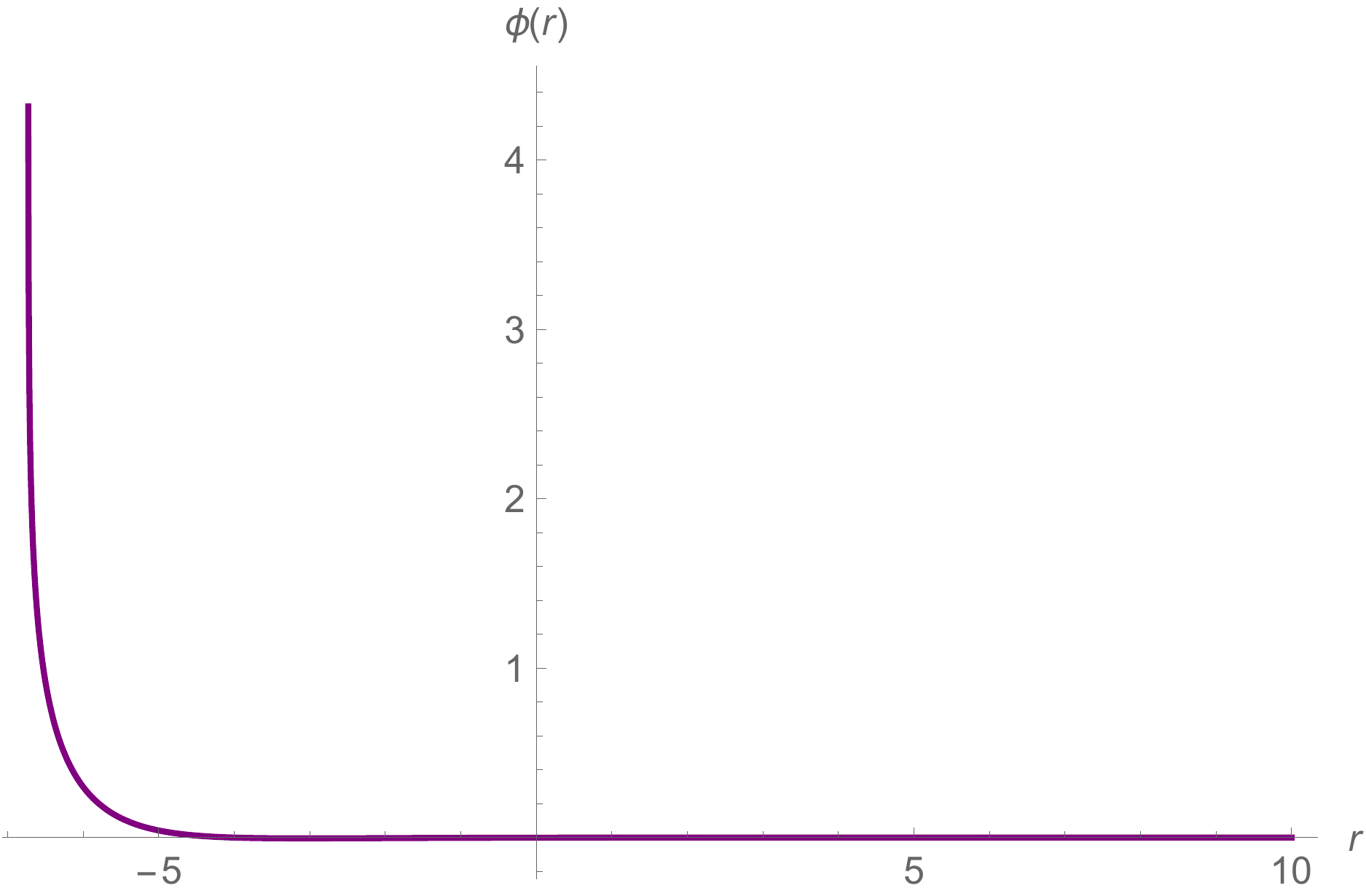}
  \caption{$\phi(r)$ solution}
  \end{subfigure}
  \begin{subfigure}[b]{0.45\linewidth}
    \includegraphics[width=\linewidth]{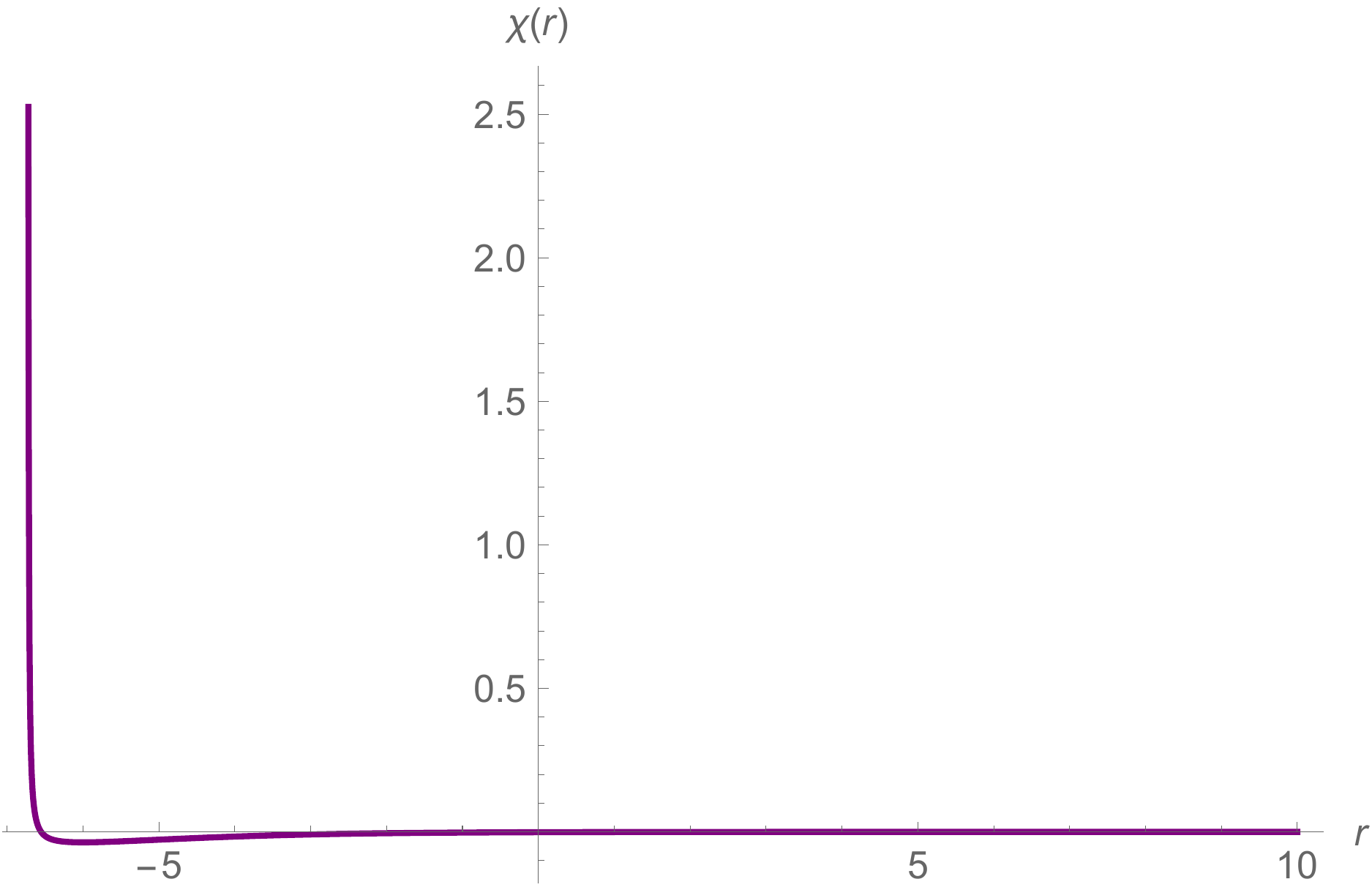}
  \caption{$\chi(r)$ solution}
  \end{subfigure}\\
  \begin{subfigure}[b]{0.45\linewidth}
    \includegraphics[width=\linewidth]{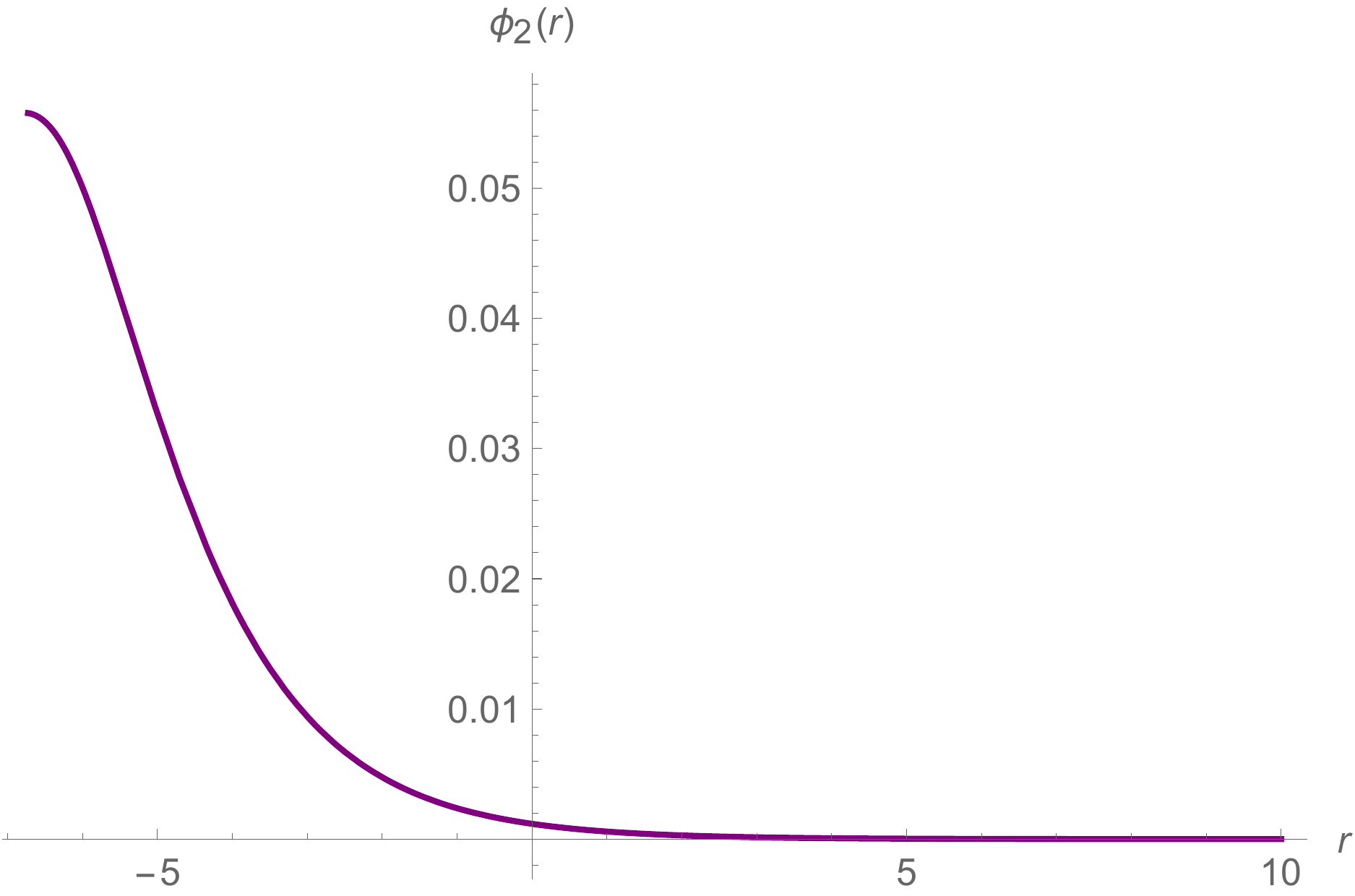}
  \caption{$\phi_2(r)$ solution}
  \end{subfigure}
  \begin{subfigure}[b]{0.45\linewidth}
    \includegraphics[width=\linewidth]{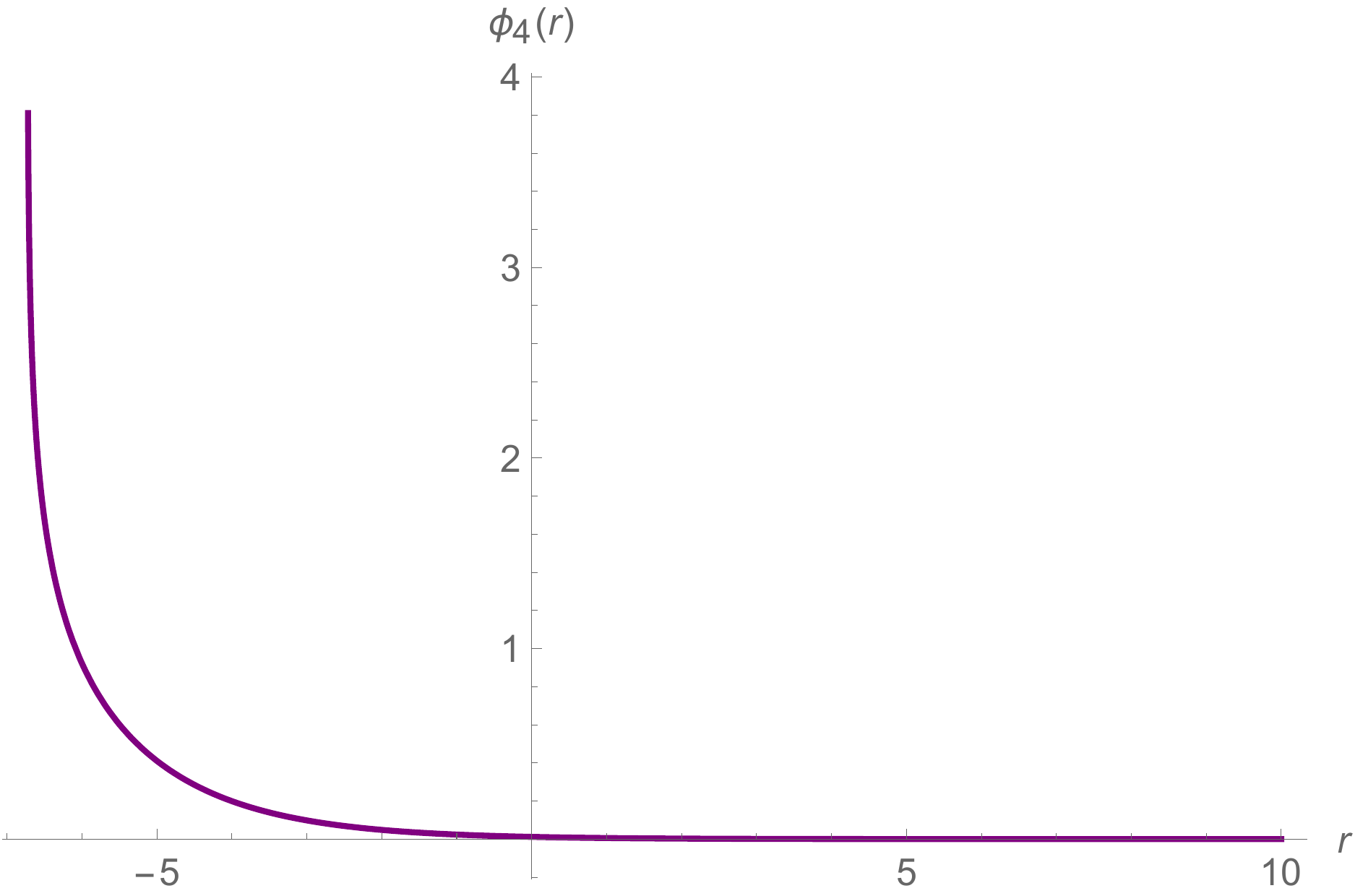}
  \caption{$\phi_4(r)$ solution}
  \end{subfigure}\\
   \begin{subfigure}[b]{0.45\linewidth}
    \includegraphics[width=\linewidth]{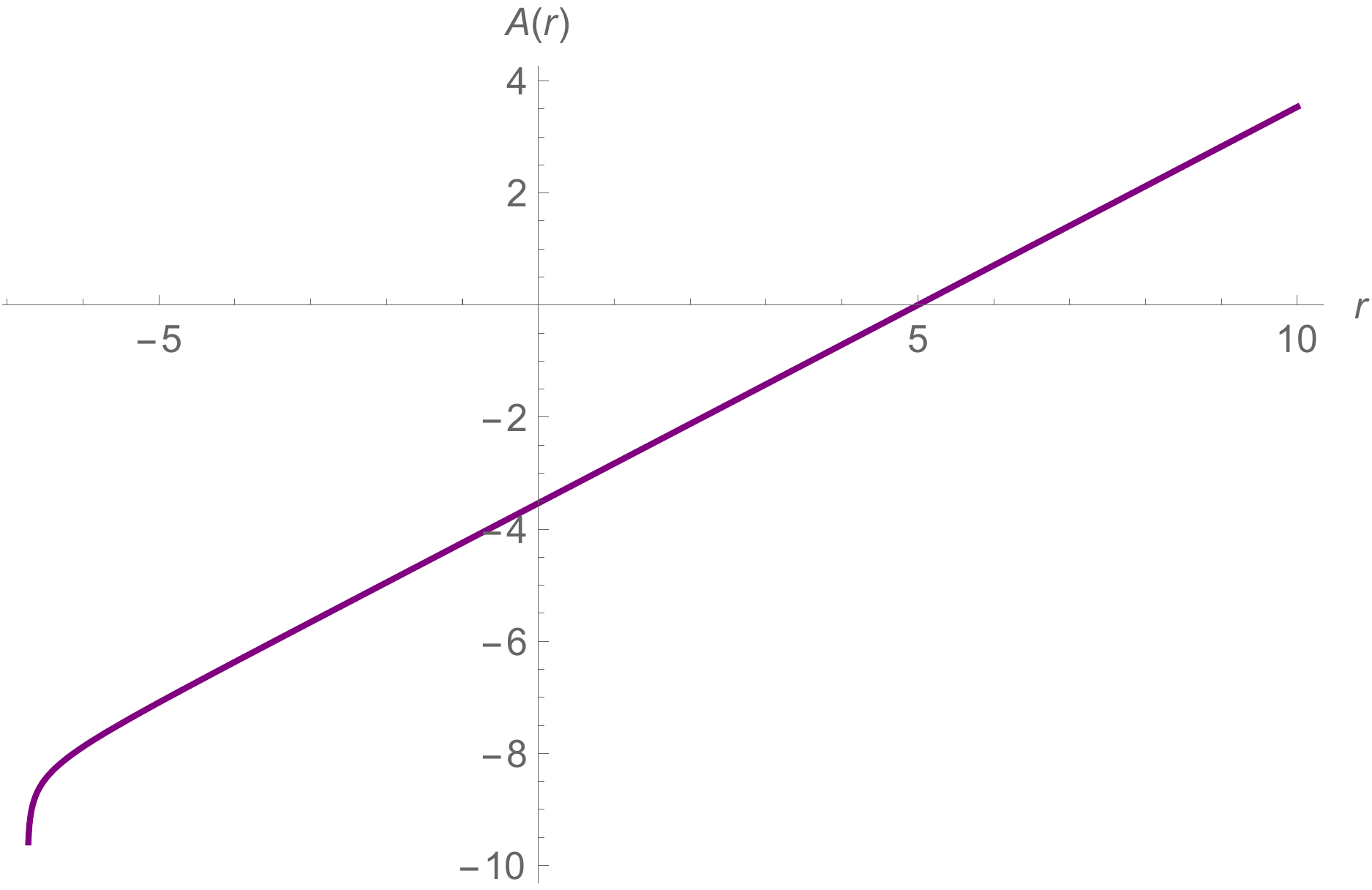}
  \caption{$A(r)$ solution}
   \end{subfigure} 
 \begin{subfigure}[b]{0.45\linewidth}
    \includegraphics[width=\linewidth]{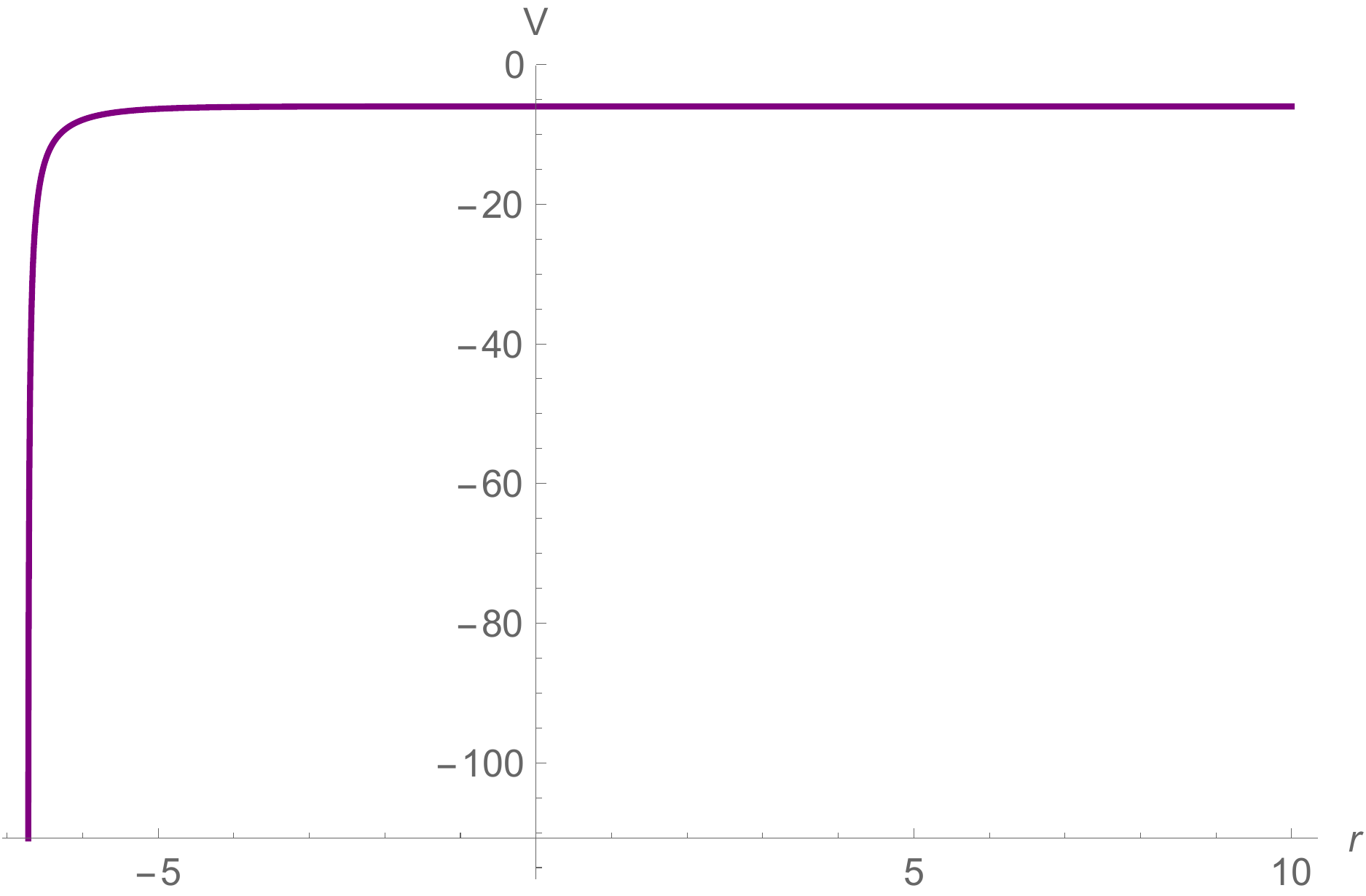}
  \caption{Scalar potential}
   \end{subfigure} 
  \caption{An $N=2$ RG flow from the $N=4$ SCFT in the UV to a non-conformal phase in the IR for $g_1=g_2=\frac{1}{\sqrt{2}}$.}
  \label{Fig1}
\end{figure}

\begin{figure}
  \centering
   \begin{subfigure}[b]{0.45\linewidth}
    \includegraphics[width=\linewidth]{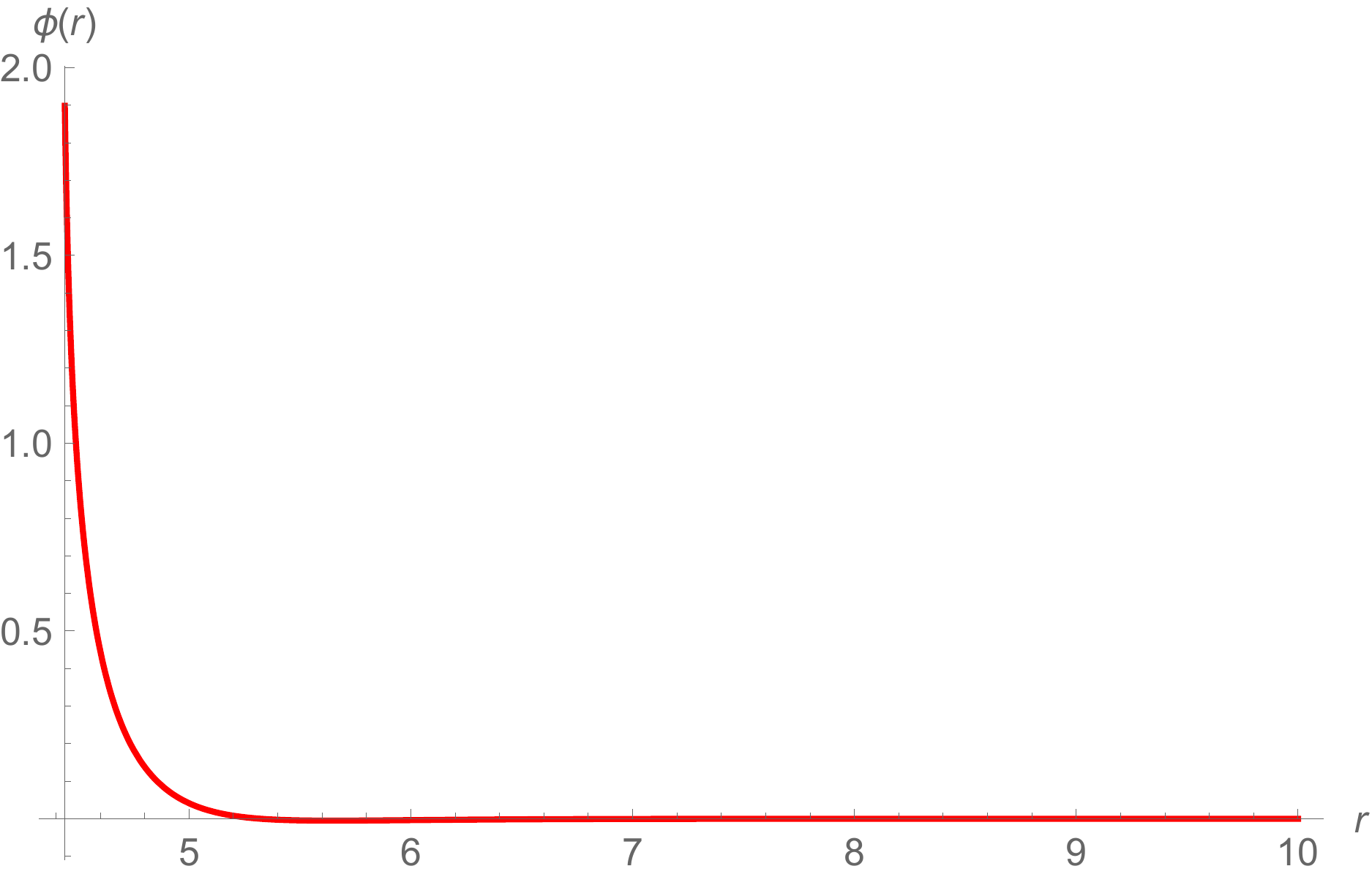}
  \caption{$\phi(r)$ solution}
  \end{subfigure}
  \begin{subfigure}[b]{0.45\linewidth}
    \includegraphics[width=\linewidth]{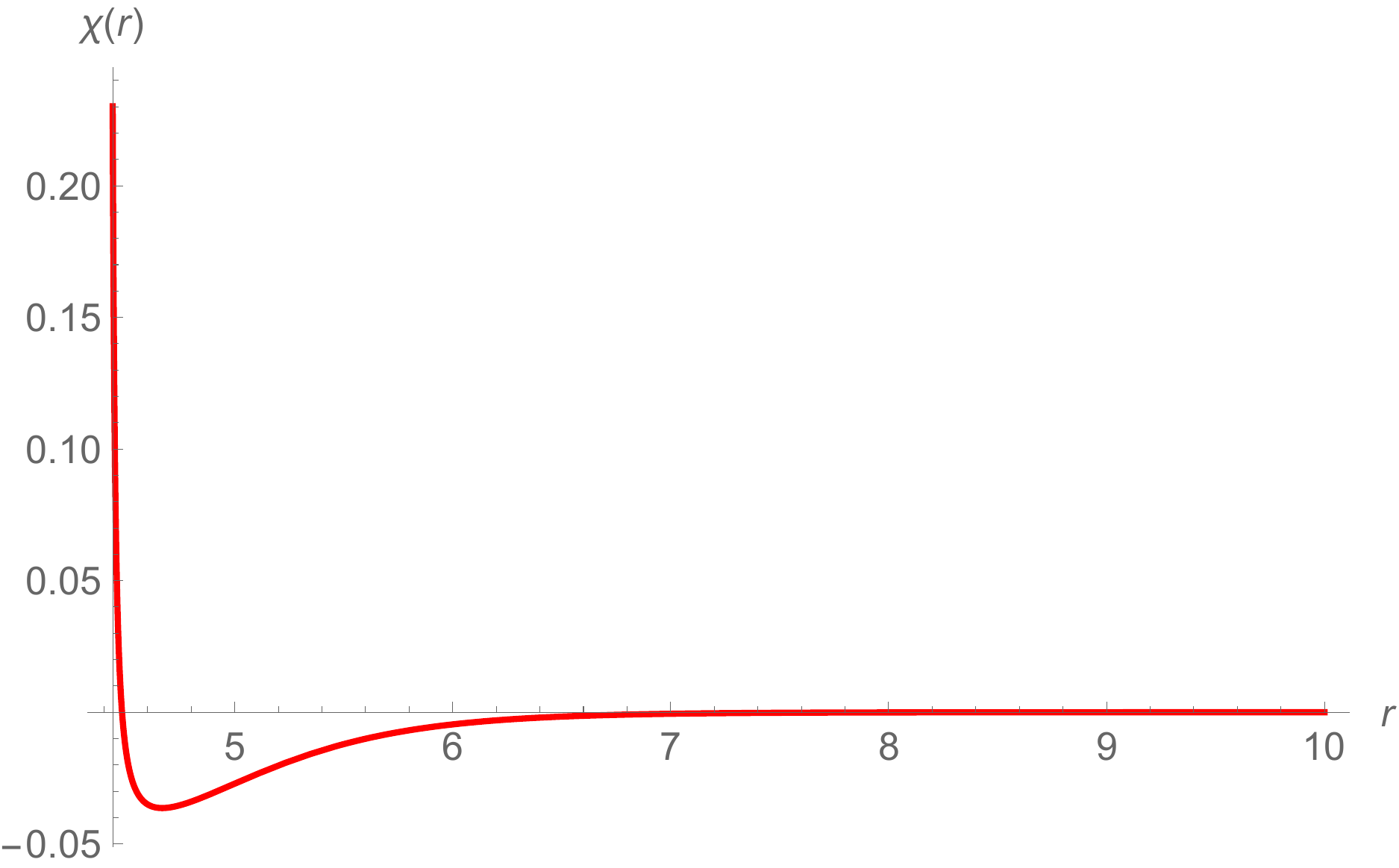}
  \caption{$\chi(r)$ solution}
  \end{subfigure}\\
  \begin{subfigure}[b]{0.45\linewidth}
    \includegraphics[width=\linewidth]{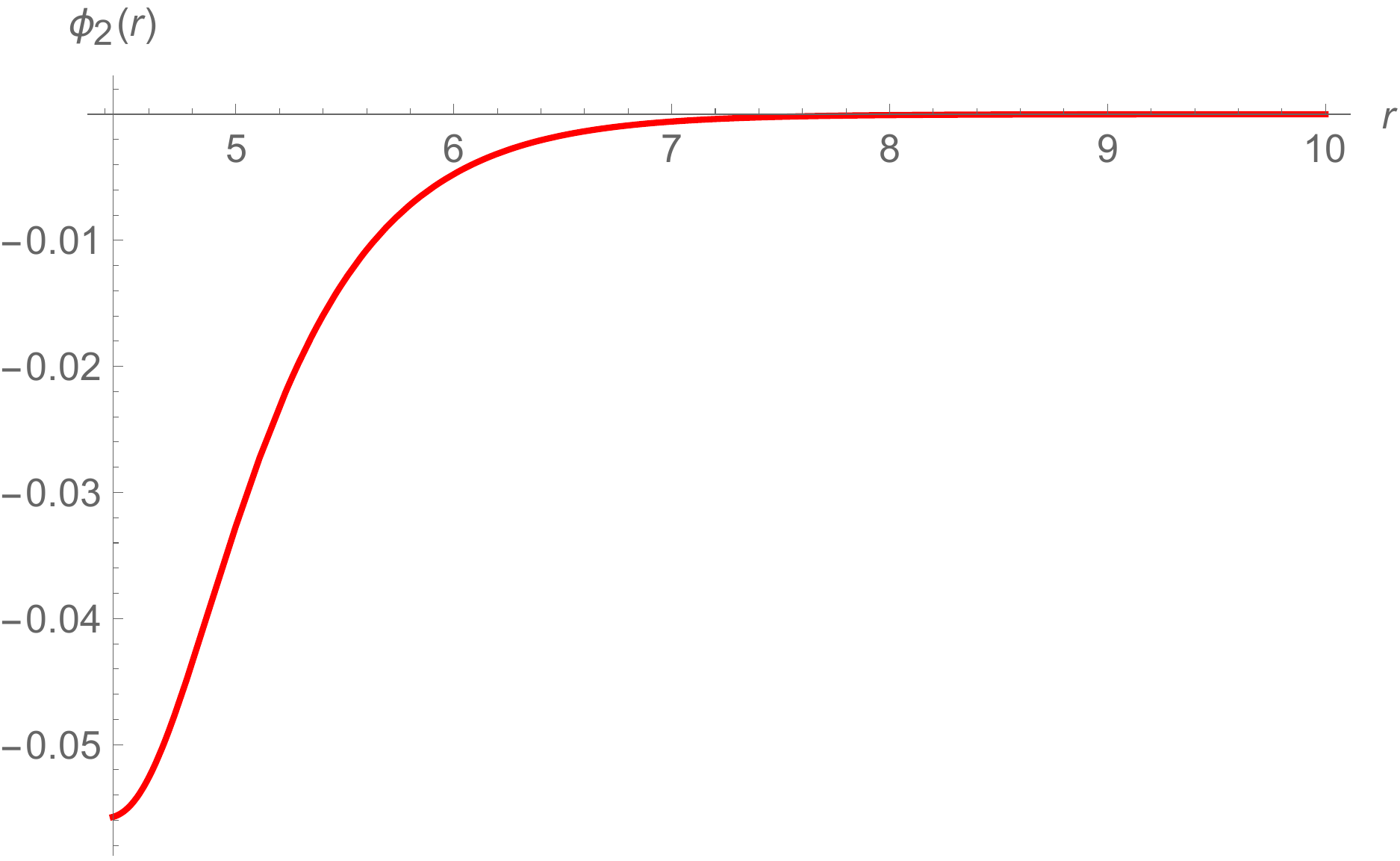}
  \caption{$\phi_2(r)$ solution}
  \end{subfigure}
  \begin{subfigure}[b]{0.45\linewidth}
    \includegraphics[width=\linewidth]{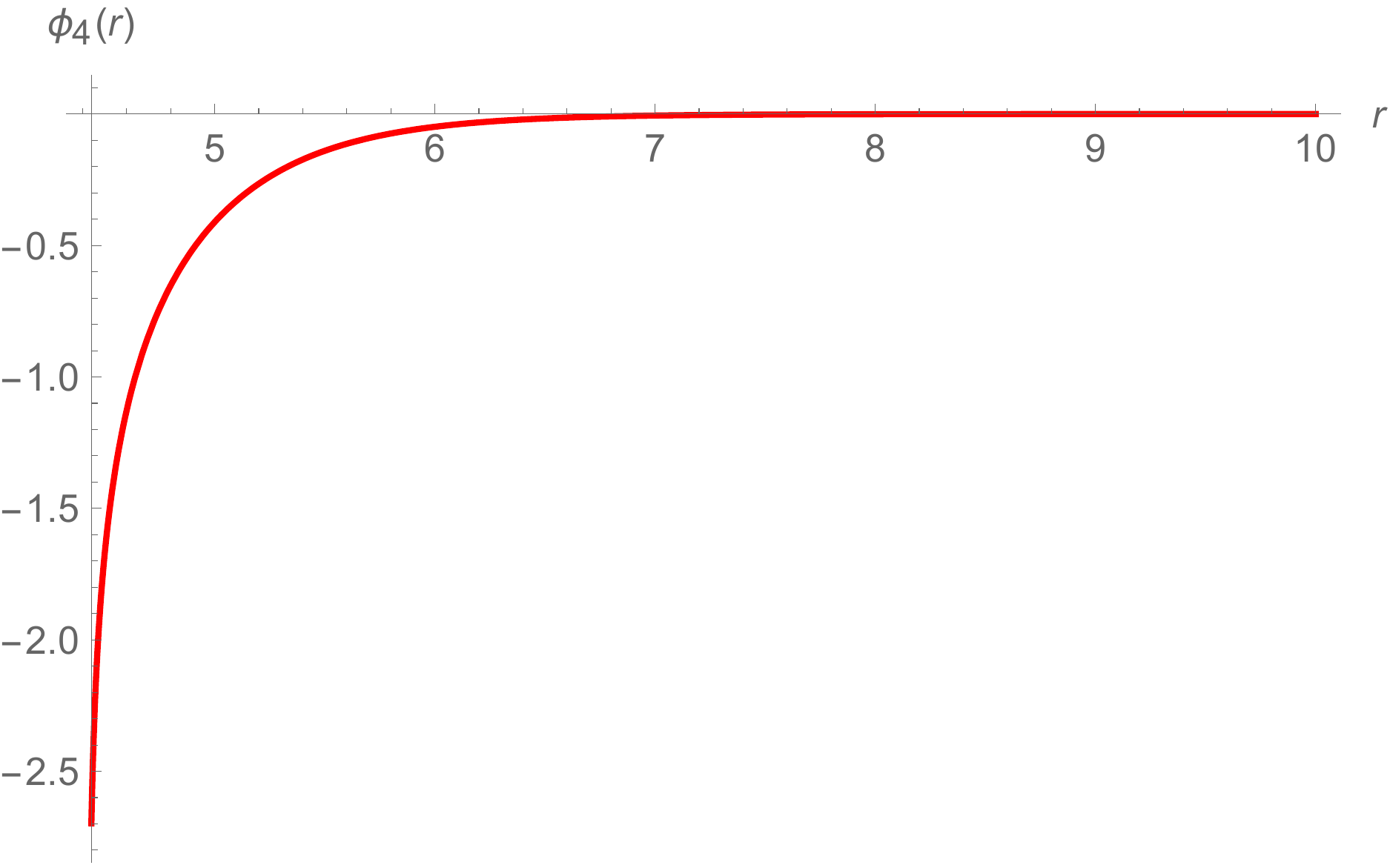}
  \caption{$\phi_4(r)$ solution}
  \end{subfigure}\\
   \begin{subfigure}[b]{0.45\linewidth}
    \includegraphics[width=\linewidth]{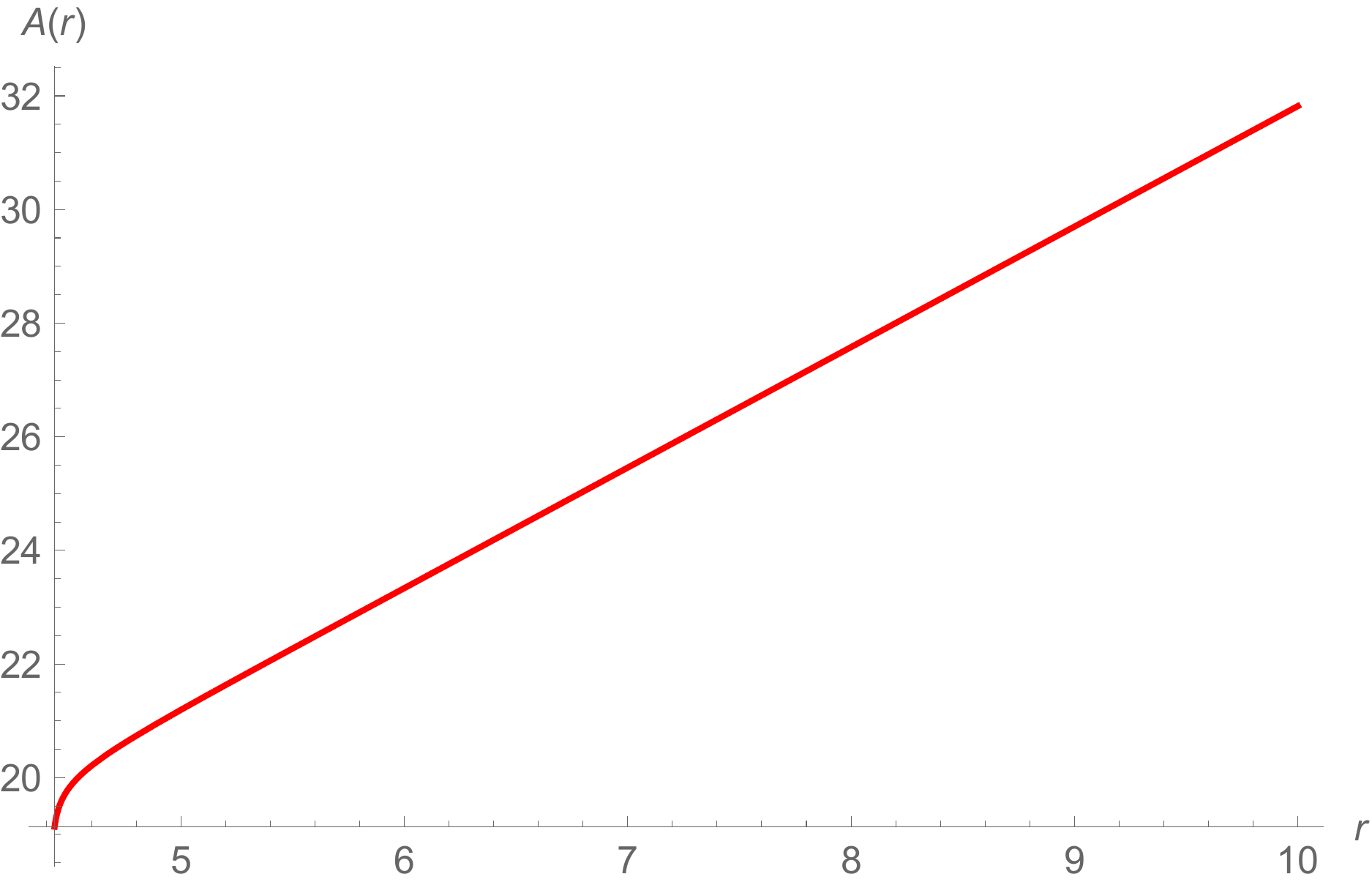}
  \caption{$A(r)$ solution}
   \end{subfigure} 
 \begin{subfigure}[b]{0.45\linewidth}
    \includegraphics[width=\linewidth]{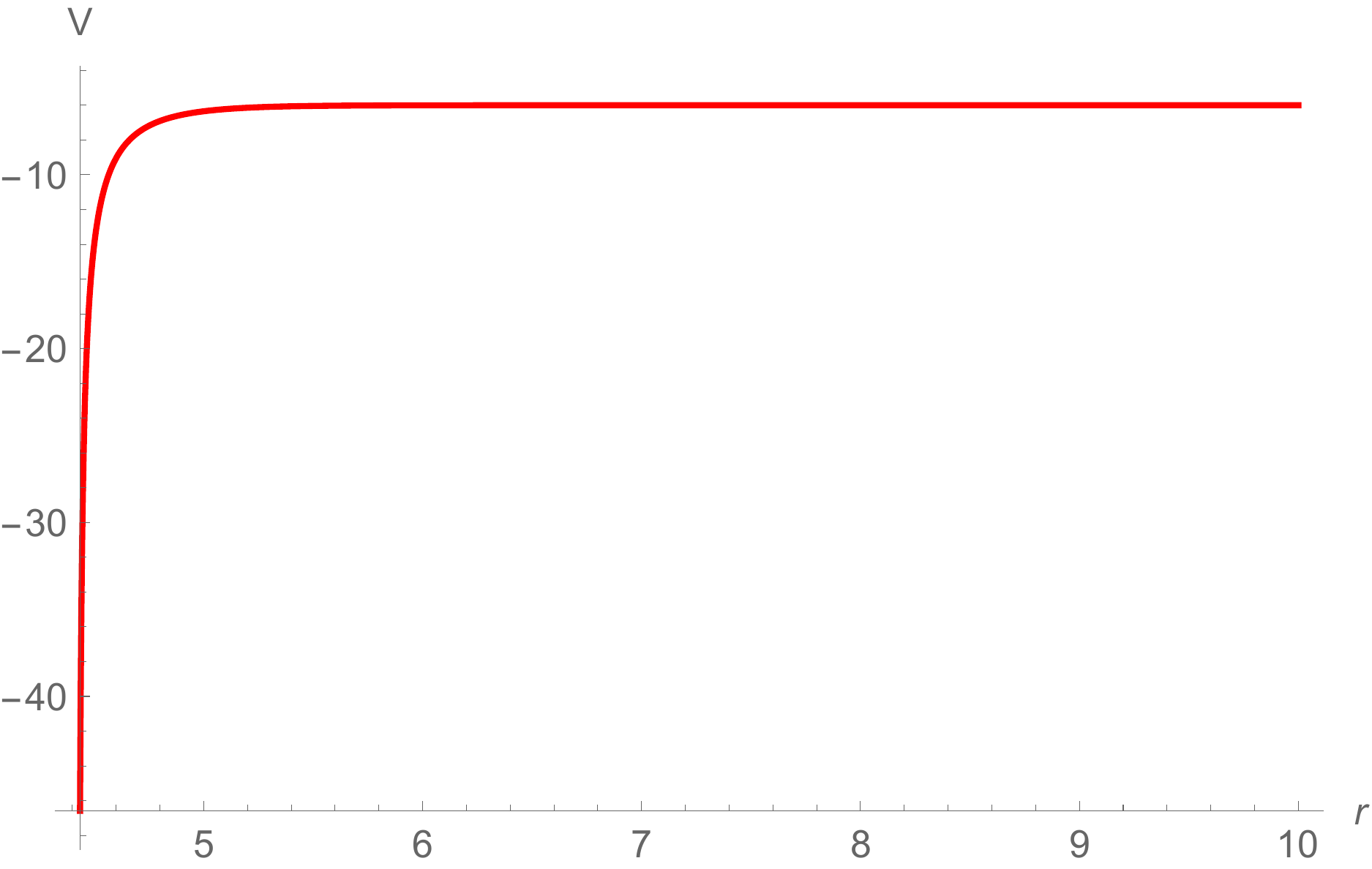}
  \caption{Scalar potential}
   \end{subfigure} 
  \caption{An $N=2$ RG flow from the $N=4$ SCFT in the UV to a non-conformal phase in the IR for $g_1=g_2=\frac{3\sqrt{2}}{2}$.}
  \label{Fig2}
\end{figure}

\begin{figure}
  \centering
   \begin{subfigure}[b]{0.45\linewidth}
    \includegraphics[width=\linewidth]{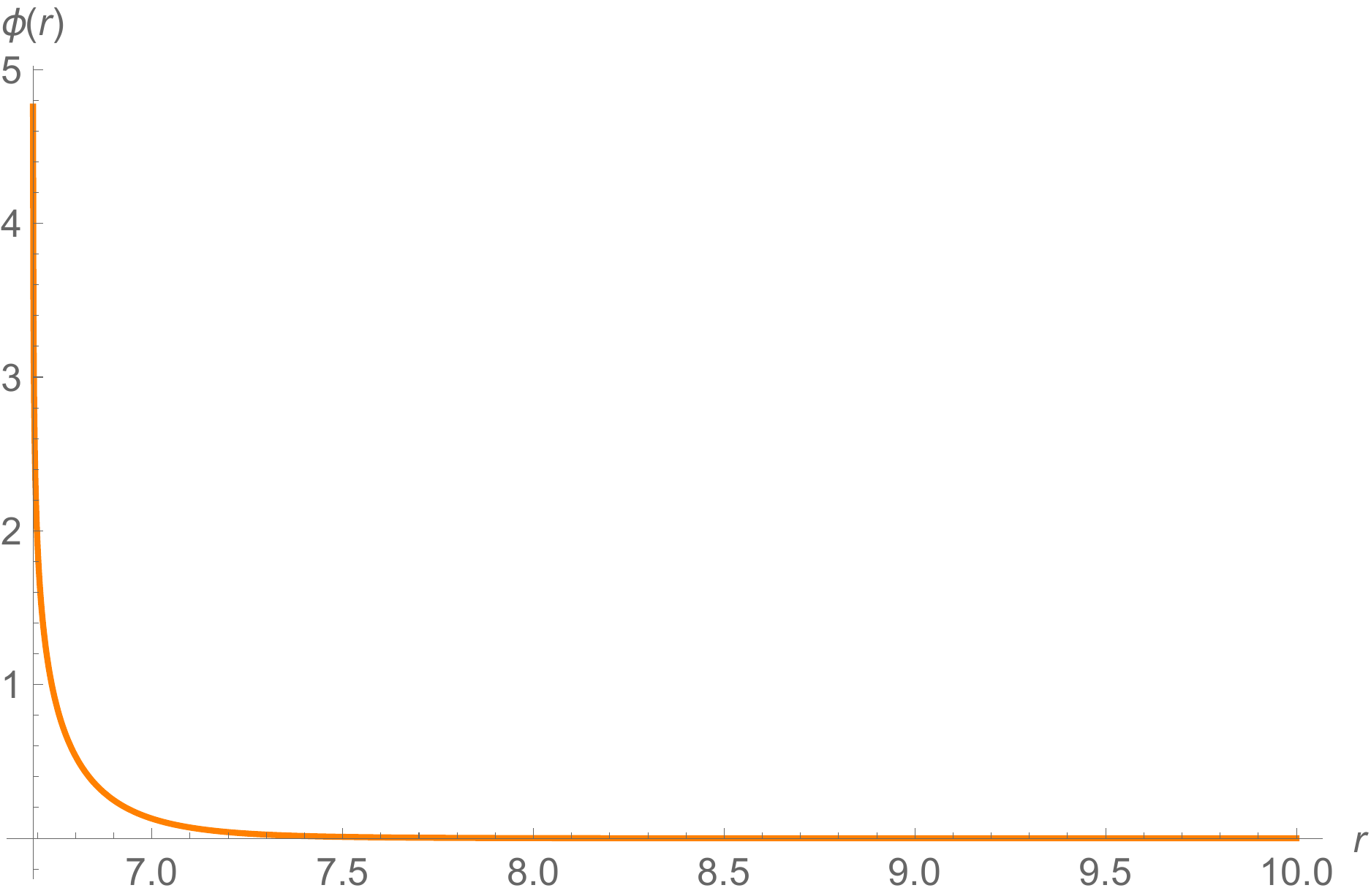}
  \caption{$\phi(r)$ solution}
  \end{subfigure}
  \begin{subfigure}[b]{0.45\linewidth}
    \includegraphics[width=\linewidth]{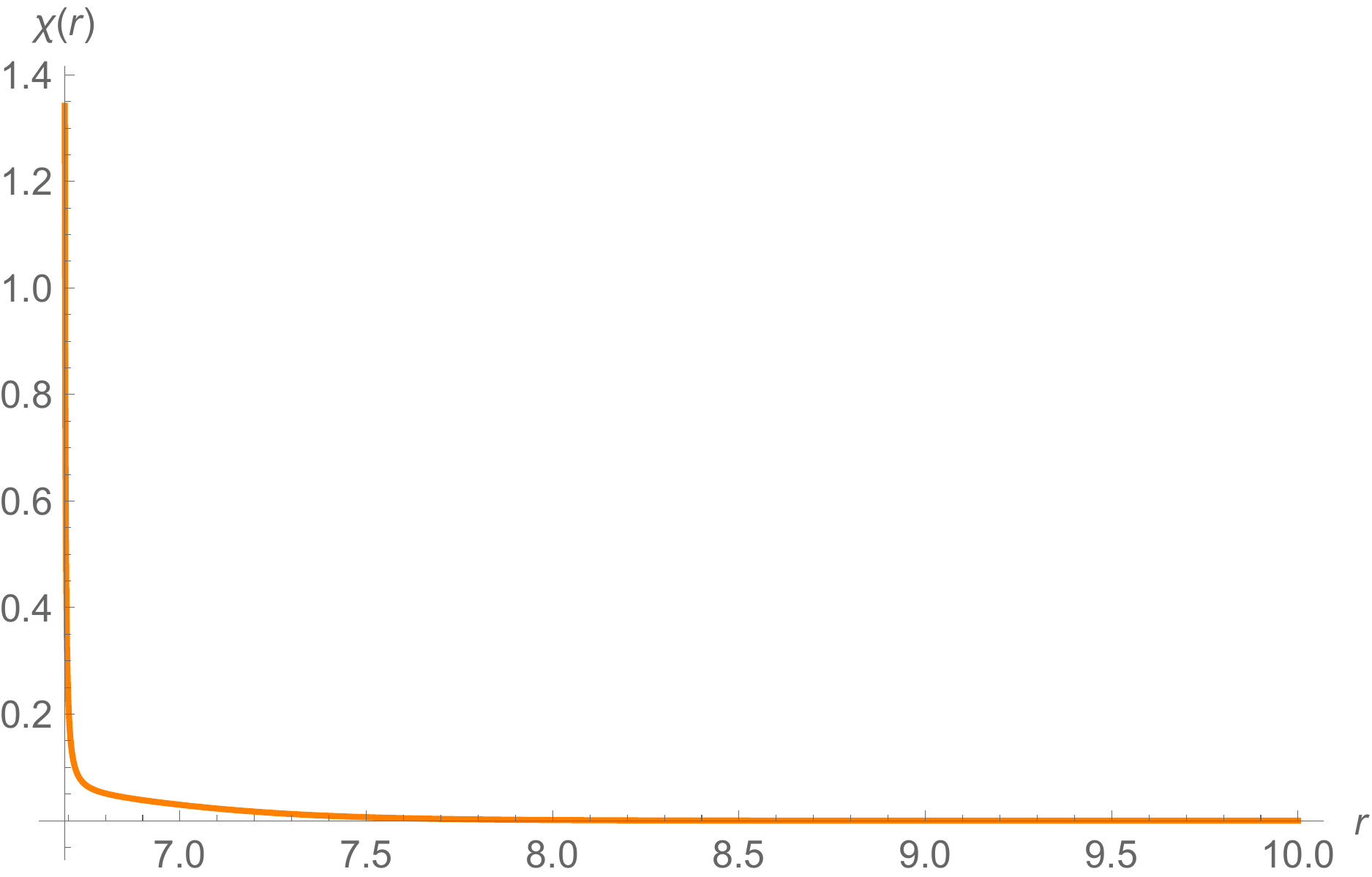}
  \caption{$\chi(r)$ solution}
  \end{subfigure}\\
  \begin{subfigure}[b]{0.45\linewidth}
    \includegraphics[width=\linewidth]{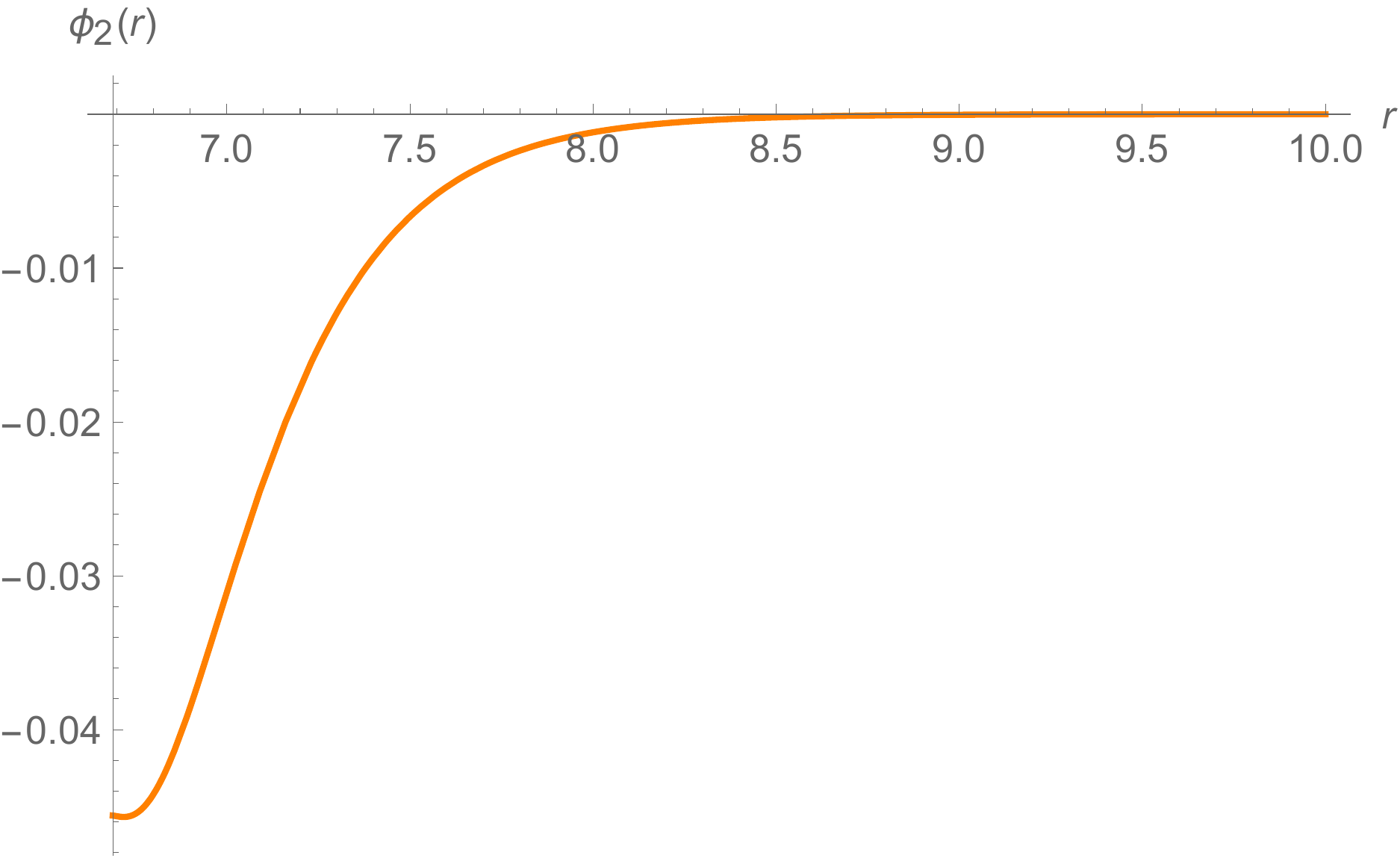}
  \caption{$\phi_2(r)$ solution}
  \end{subfigure}
  \begin{subfigure}[b]{0.45\linewidth}
    \includegraphics[width=\linewidth]{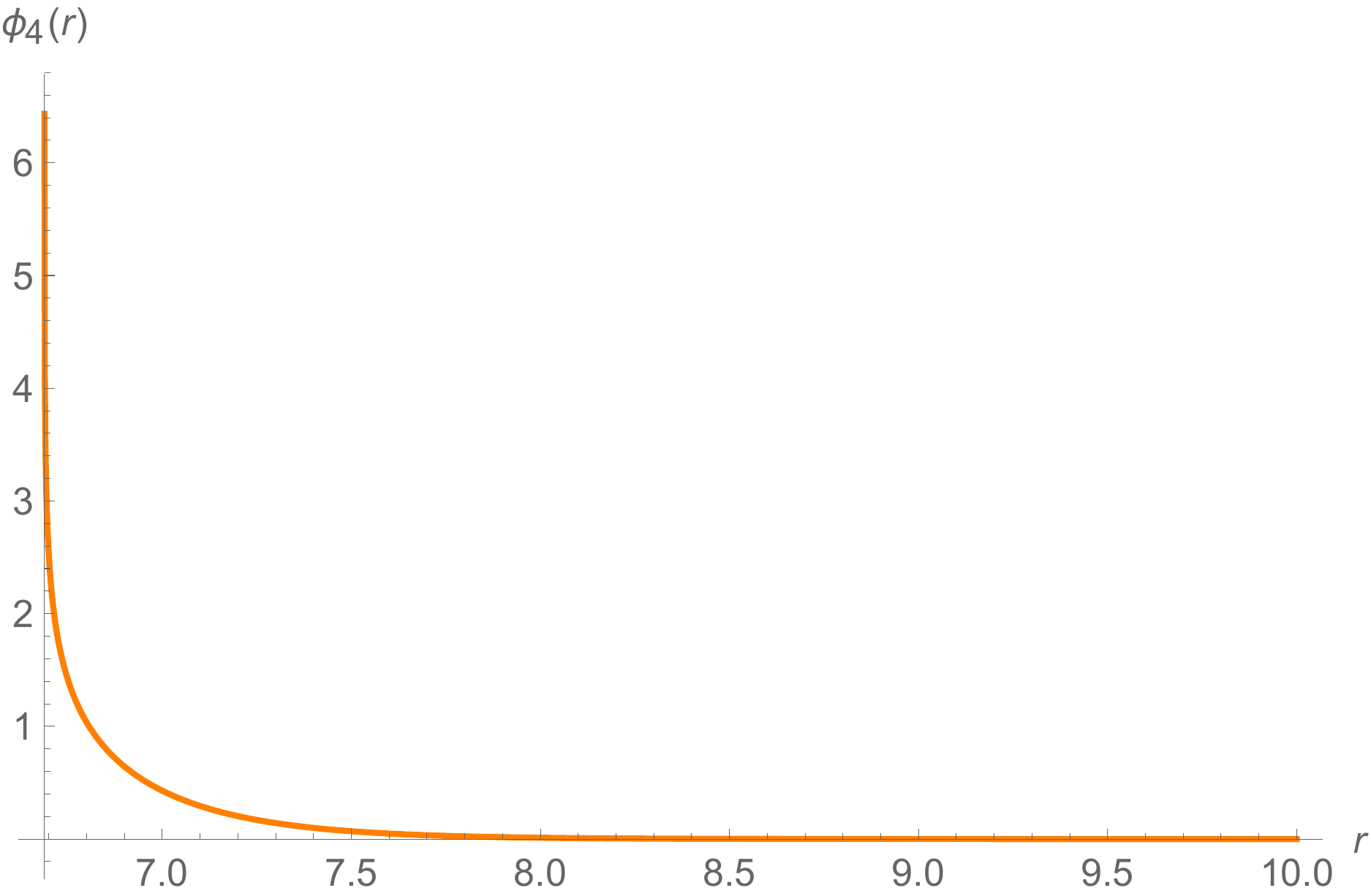}
  \caption{$\phi_4(r)$ solution}
  \end{subfigure}\\
   \begin{subfigure}[b]{0.45\linewidth}
    \includegraphics[width=\linewidth]{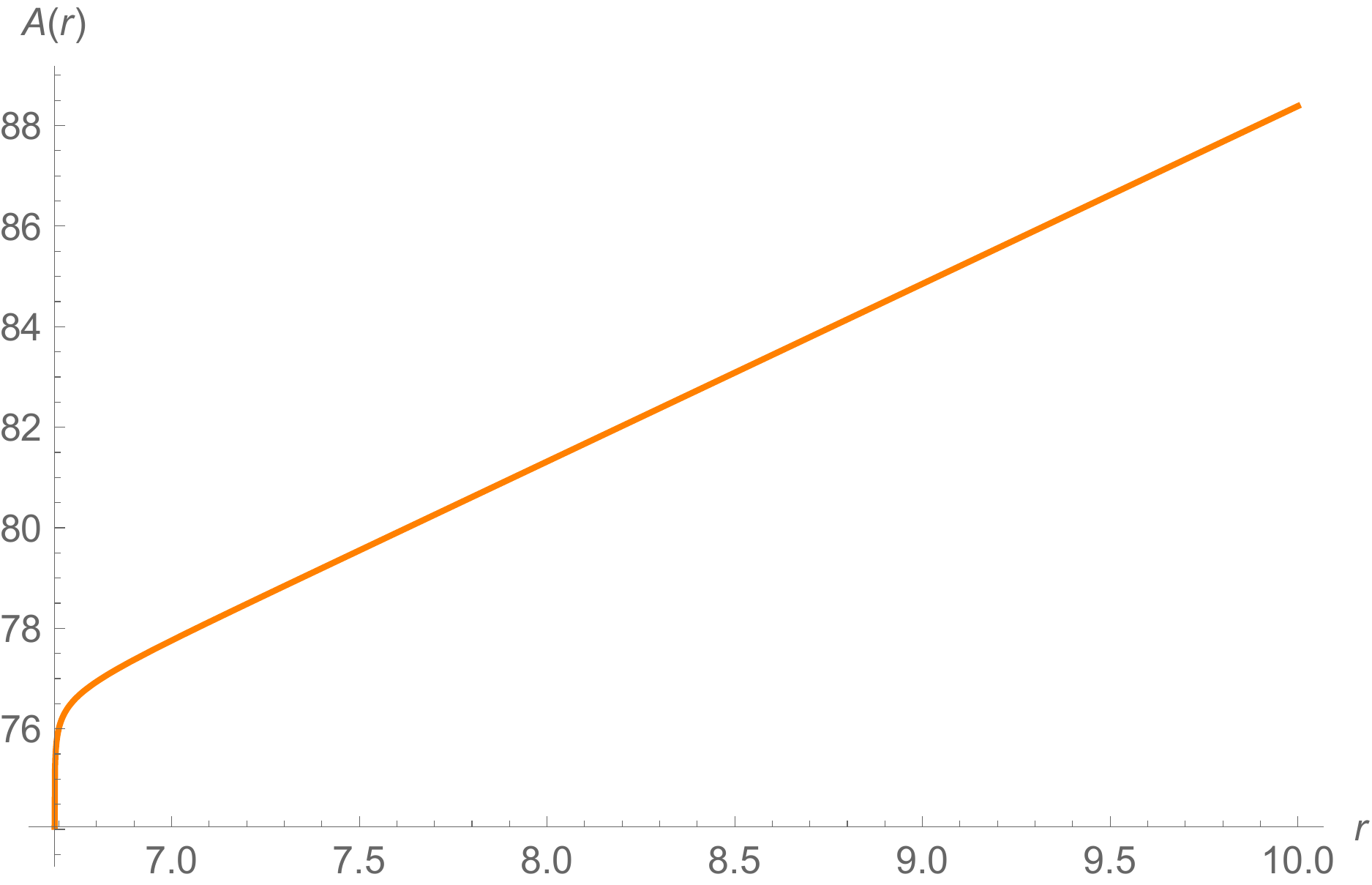}
  \caption{$A(r)$ solution}
   \end{subfigure} 
 \begin{subfigure}[b]{0.45\linewidth}
    \includegraphics[width=\linewidth]{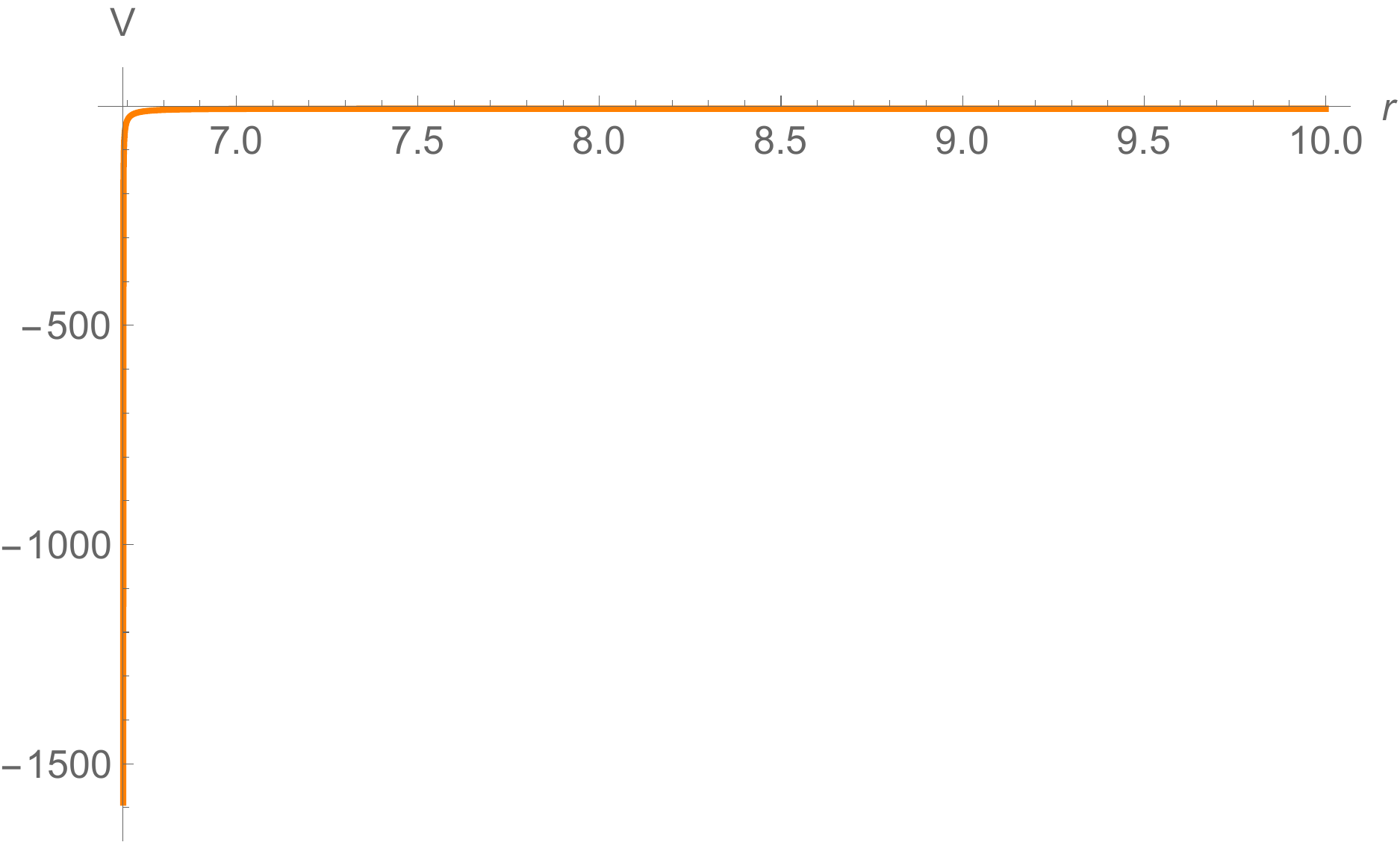}
  \caption{Scalar potential}
   \end{subfigure} 
  \caption{An $N=2$ RG flow from the $N=4$ SCFT in the UV to a non-conformal phase in the IR for $g_1=g_2=\frac{5\sqrt{2}}{2}$.}
  \label{Fig3}
\end{figure}

We now consider the most general solutions in the $SO(2)\times SO(2)\times SO(2)\times SO(2)$ truncation with non-vanishing $\chi$ and $\phi_1=\phi_3=0$. In this case, the $SO(4)\subset SO(6)_R$ is broken to $SO(2)\times SO(2)$ while the $SO(4)$ subgroup of the $SO(6)$ symmetry of the vector multiplets is broken to $SO(3)\times SO(2)$, or $SO(2)\times SO(3)$ if we choose $\phi_2=\phi_4=0$ rather than $\phi_1=\phi_3=0$.
\\
\indent With $\phi_1=\phi_3=0$, the superpotential is given by
\begin{equation}
\mc{W}=\mc{W}_+=\frac{1}{2}\left[g_1e^{-\frac{\phi}{2}}(\cosh\phi_2+i\sinh\phi_2\sinh\phi_4)+g_2e^{\frac{\phi}{2}}\cosh\phi_4+ig_2e^{-\frac{\phi}{2}}\chi\cosh\phi_4\right]\label{W_N2_gen}
\end{equation}  
in which for definiteness, we have chosen $\mc{W}_+$ to be the superpotential with $\epsilon^{1}$ and $\epsilon^4$ being the corresponding Killing spinors. If we choose $\mc{W}=\mc{W}_-$, the unbroken supersymmetry will correspond to $\epsilon^2$ and $\epsilon^3$.
\\
\indent With the same analysis as in the previous cases, we find the following BPS equations
\begin{eqnarray}
A'&=&W\nonumber \\
&=&\frac{1}{2}\sqrt{(g_1e^{-\frac{\phi}{2}}\cosh\phi_2+g_2e^{\frac{\phi}{2}}\cosh\phi_4)^2+e^{-\phi}(g_1\sinh\phi_2\sinh\phi_4+g_2\cosh\phi_4\chi)^2},\nonumber \\
& &\\
\phi'&=&-4\frac{\pd W}{\pd\phi}\nonumber \\
&=&\frac{1}{8W}e^{-\phi}\left[g_1^2(3+\cosh2\phi_2+2\cosh2\phi_4\sinh^2\phi_2)-4g_2^2e^{2\phi}\cosh^2\phi_4 \right.\nonumber \\
& &\left. +4g_2\chi(g_1\sinh\phi_2\sinh2\phi_4+g_2\chi\cosh^2\phi_4)\right],\\
\chi'&=&-4e^{2\phi}\frac{\pd W}{\pd \chi}=-\frac{e^\phi}{W}g_2\cosh\phi_4(g_1\sinh\phi_2\sinh\phi_4+g_2\cosh\phi_4\chi),\\
\phi_2'&=&-\frac{2}{\cosh^2\phi_4}\frac{\pd W}{\pd \phi_2}\nonumber \\
&=&-\frac{1}{2W}e^{-\phi}g_1\left[\sinh\phi_2(g_1\cosh\phi_2+g_2e^\phi\textrm{sech}\phi_4) +g_2\chi \cosh\phi_2\tanh\phi_4 \right],\quad \,\,\, \\
\phi_4'&=&-2\frac{\pd W}{\pd \phi_4}\nonumber \\
&=&-\frac{1}{4W}e^{-\phi}\left[\sinh2\phi_4(g_2^2e^{2\phi}+g_2^2\chi^2+g_1^2\sinh^2\phi_2)\ \right. \nonumber \\
& &\left.+2g_1g_2e^\phi\cosh\phi_2\sinh\phi_4+2g_1g_2\chi\cosh2\phi_4\sinh\phi_2\right].
\end{eqnarray}
\indent In this case, we are not able to find analytic solutions, so we will instead perform a numerical analysis for solving these equations. By choosing the numerical values of the coupling constants $g_1=g_2=\sqrt{2}g_0$, we obtain examples of numerical RG flow solutions for suitable boundary conditions and different values of the parameter $g_0$ as shown in figures \ref{Fig1}, \ref{Fig2} and \ref{Fig3}. From these figures, we see some patterns namely, in all of these solutions, $\phi_2$ is constant near the singularities as in the previous cases, and all the singularities are physically acceptable. The former indeed seems to be a general feature at least among all the solutions we have found. The latter follows directly from the scalar potential \eqref{Potential} for $g_2=g_1$, 
\begin{equation}
V=-\frac{1}{2}g_1^2e^{-\phi}(1+e^{2\phi}+\chi^2+4e^{\phi}\cosh\phi_2\cosh\phi_4)
\end{equation}
from which we immediately see that $V\rightarrow -\infty$ for any diverging behaviors of all the scalar fields. This implies that all possible singularities in this $SO(2)\times SO(2)\times SO(3)\times SO(2)$ truncation are physical. Accordingly, the corresponding solutions describe RG flows from the dual $N=4$ SCFT to various non-conformal phases in the IR.

\section{Supersymmetric Janus solutions}\label{Janus}
In this section, we look at supersymmetric Janus solutions with $N=4$ and $N=2$ supersymmetries. In this case, the resulting BPS equations are much more complicated than those in the RG flow case, and we are not able to obtain any analytic solutions.   
   
\subsection{$N=4$ Janus solutions with $SO(4)\times SO(4)$ symmetry}   
\begin{figure}
  \centering
  \begin{subfigure}[b]{0.5\linewidth}
    \includegraphics[width=\linewidth]{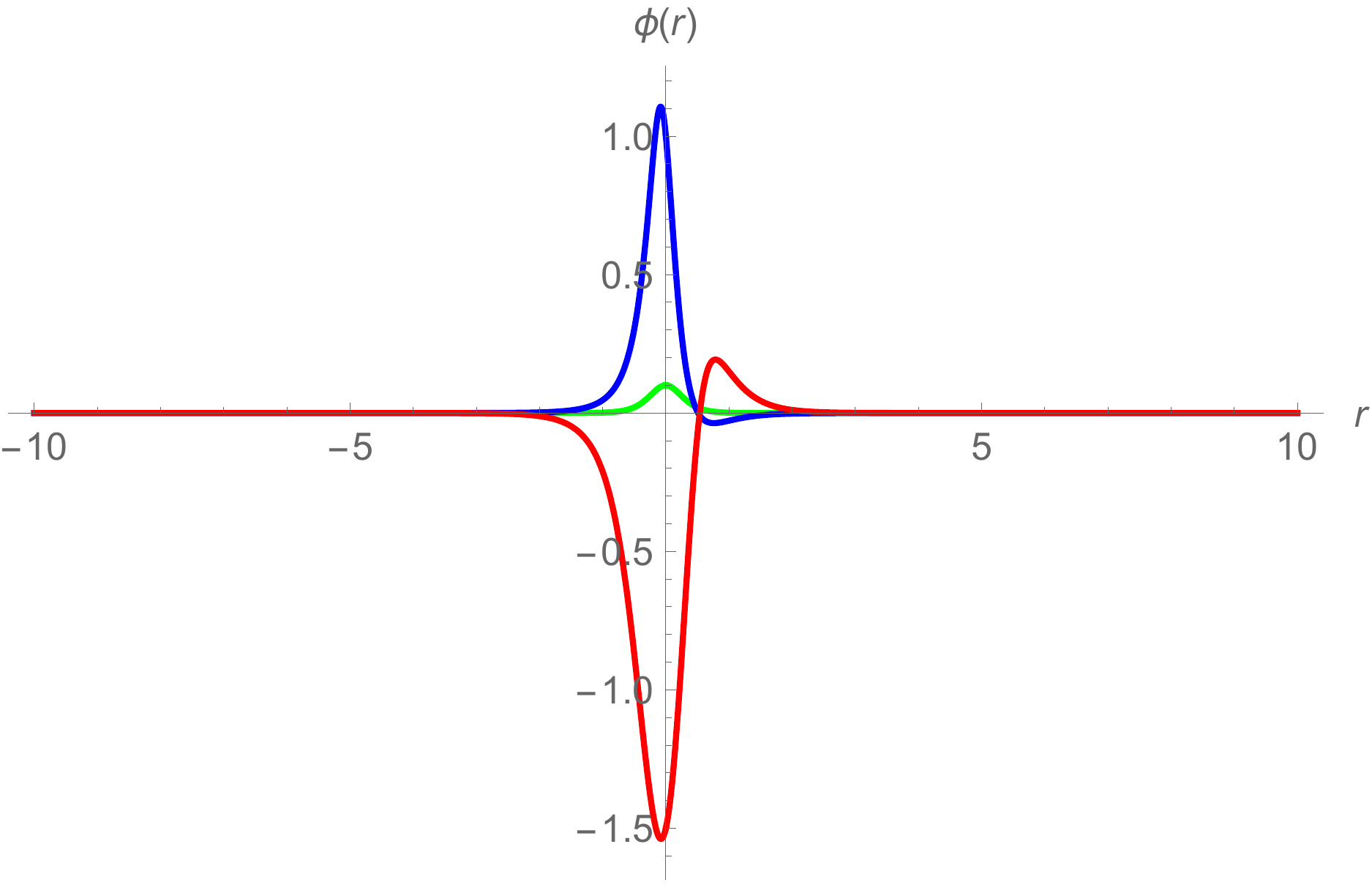}
  \caption{$\phi(r)$ solutions}
  \end{subfigure}\\
  \begin{subfigure}[b]{0.5\linewidth}
    \includegraphics[width=\linewidth]{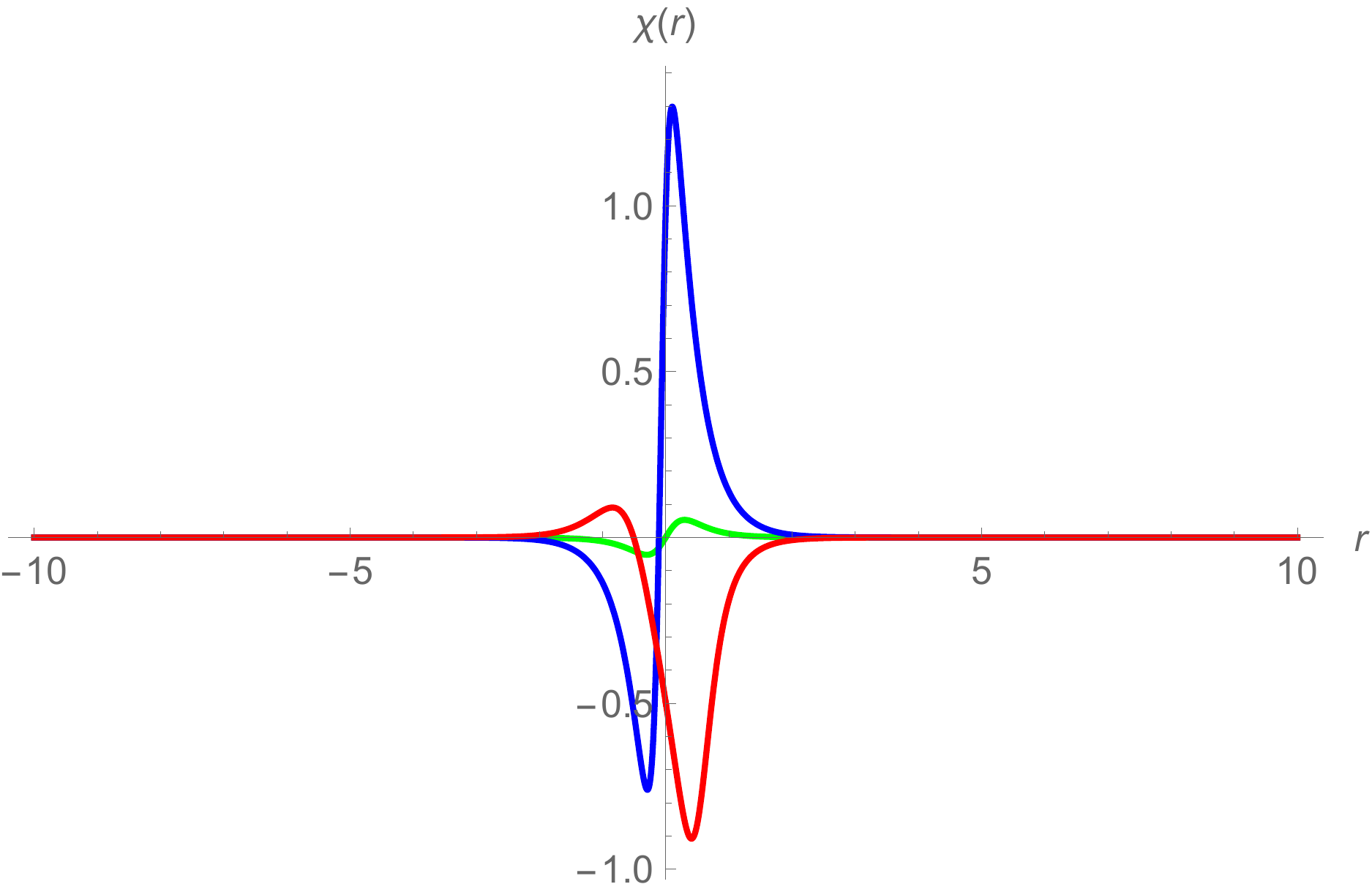}
  \caption{$\chi(r)$ solutions}
  \end{subfigure}\\
  \begin{subfigure}[b]{0.5\linewidth}
    \includegraphics[width=\linewidth]{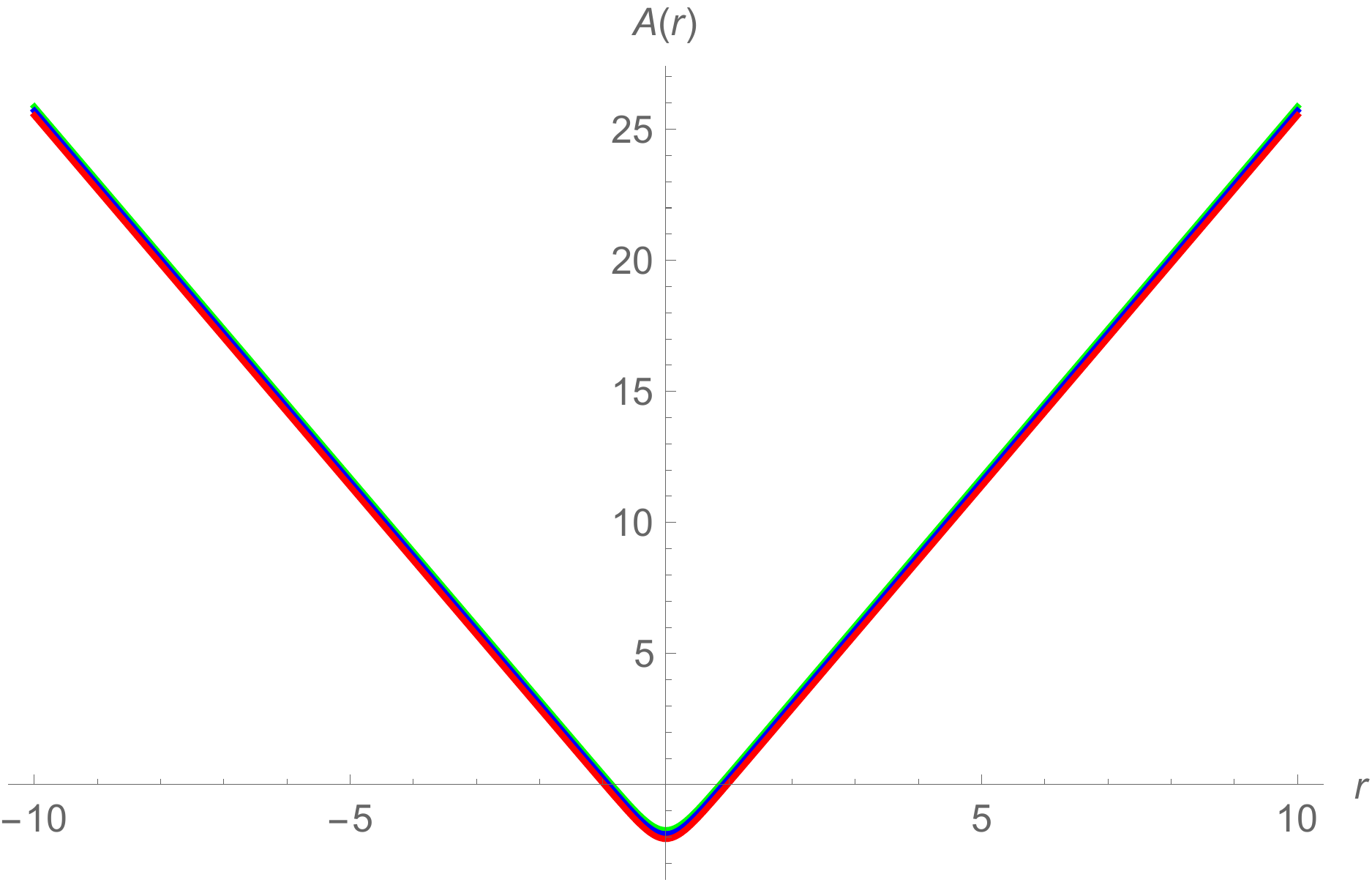}
  \caption{$A(r)$ solutions}
  \end{subfigure}
  \caption{$N=4$ supersymmetric Janus solutions from $N=4$ $SO(4)\times SO(4)$ gauged supergravity for $g_1=g_2=2\sqrt{2}$ and $\ell=\kappa=1$ with different boundary conditions.}
  \label{FigJ1}
\end{figure}

We first consider solutions with all $\phi_i=0$ that can be regarded as solutions of pure $N=4$ gauged supergravity with $SO(4)\sim SO(3)\times SO(3)$ gauge group. In this case, we find a simple superpotential
\begin{equation}
\mc{W}=\frac{1}{2}e^{-\frac{\phi}{2}}[g_1+g_2(e^\phi+i\chi)].
\end{equation}
Using the prescription given in section \ref{N4_SUGRA}, we obtain the following BPS equations
\begin{eqnarray}
\phi'&=&-4\frac{A'}{W}\frac{\pd W}{\pd \phi}-4\kappa \frac{e^{-A}}{\ell W}e^\phi\frac{\pd W}{\pd \chi}\nonumber \\
&=&\frac{2\ell A'(g_1^2-g_2^2e^{2\phi}+g_2^2\chi^2)-4\kappa g_2^2\chi e^{\phi-A}}{\ell[(g_1+g_2e^\phi)^2+g_2^2\chi^2]},\\
\chi'&=&-4e^{2\phi}\frac{A'}{W}\frac{\pd W}{\pd \chi}+4\kappa \frac{e^{-A}}{\ell W}e^\phi\frac{\pd W}{\pd \phi}\nonumber \\
&=&-\frac{4\ell g_2^2 A'\chi e^{2\phi}-2\kappa e^{\phi-A}(g_1^2-g_2^2e^{2\phi}+g_2^2\chi^2)}{\ell[(g_1+g_2e^\phi)^2+g_2^2\chi^2]},\\
A'^2&=&\frac{1}{4}e^{-\phi}[(g_1+g_2e^\phi)^2+g_2^2\chi^2]-\frac{e^{-2A}}{\ell^2}\, .
 \end{eqnarray} 
As expected, in the limit $\ell\rightarrow \infty$, these equations reduce to those of the RG flows studied in \cite{4D_N4_flows} after a redefinition of the scalars and coupling constants. It should also be noted that setting $\chi=0$ is not consistent unless $\ell \rightarrow \infty$. Therefore, Janus solutions are possible only for non-vanishing pseudoscalars needed to support the curvature of the $AdS_3$-sliced world-volume. This has also been pointed out in \cite{warner_Janus}.
\\
\indent We now perform a numerical analysis and give some examples of supersymmetric Janus solutions as shown in figure \ref{FigJ1}. These solutions preserve half of the original $N=4$ supersymmetry, or eight supercharges, with all $\epsilon^i$ non-vanishing and subject to the projector \eqref{gamma_r_pro}. On the conformal interfaces, the Killing spinors can have left or right chiralities depending on whether the values of $\kappa=1$ or $\kappa=-1$. The two-dimensional interfaces then preserve $(4,0)$ or $(0,4)$ supersymmetries. In the numerical analysis, we have conventionally set $\kappa=1$.

\subsection{$N=2$ Janus solutions with $SO(2)\times SO(2)\times SO(3)\times SO(2)$ symmetry}      
\begin{figure}
  \centering
  \begin{subfigure}[b]{0.45\linewidth}
    \includegraphics[width=\linewidth]{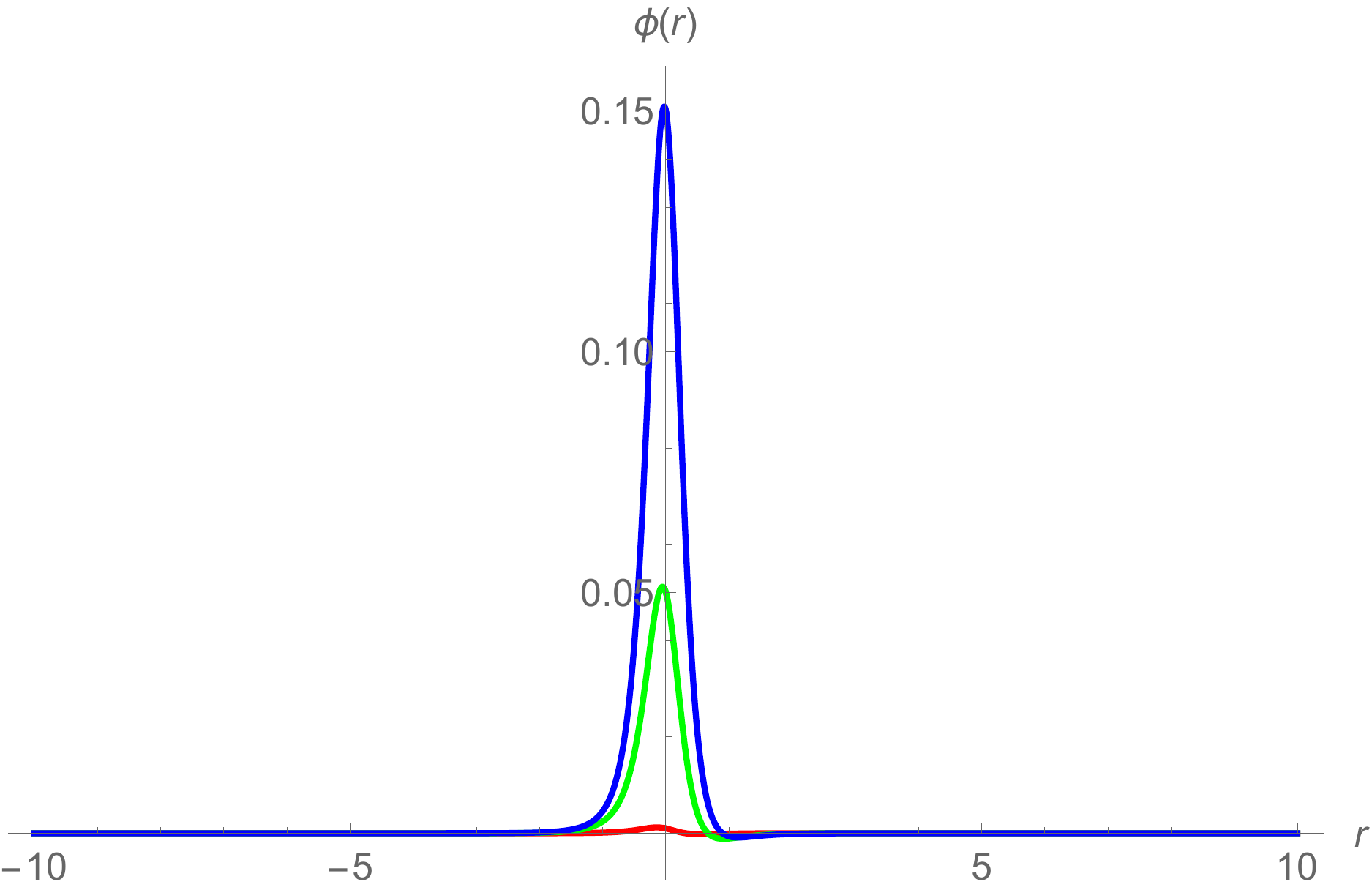}
  \caption{$\phi(r)$ solutions}
  \end{subfigure}
  \begin{subfigure}[b]{0.45\linewidth}
    \includegraphics[width=\linewidth]{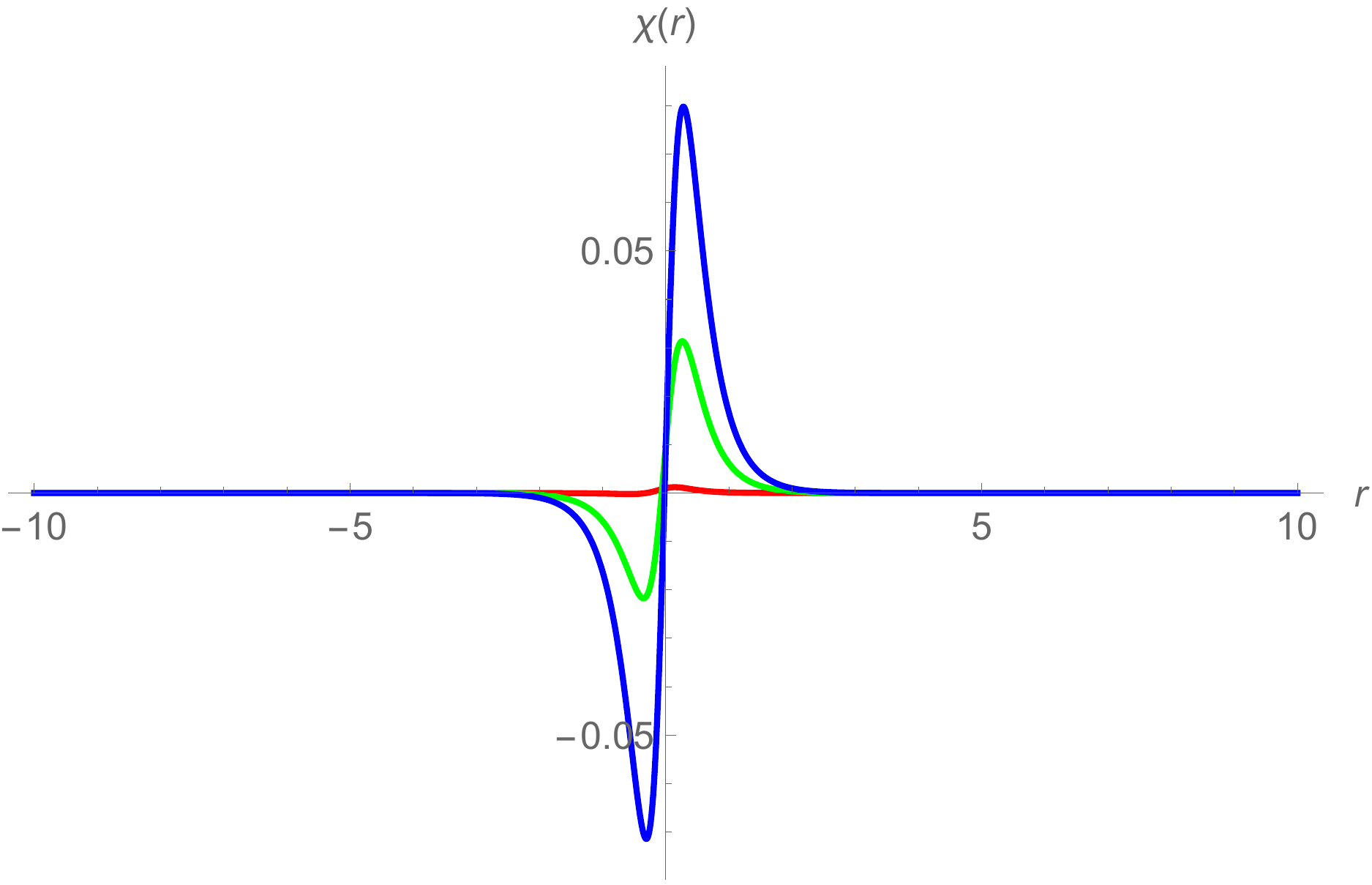}
  \caption{$\chi(r)$ solutions}
  \end{subfigure}\\
  \begin{subfigure}[b]{0.45\linewidth}
    \includegraphics[width=\linewidth]{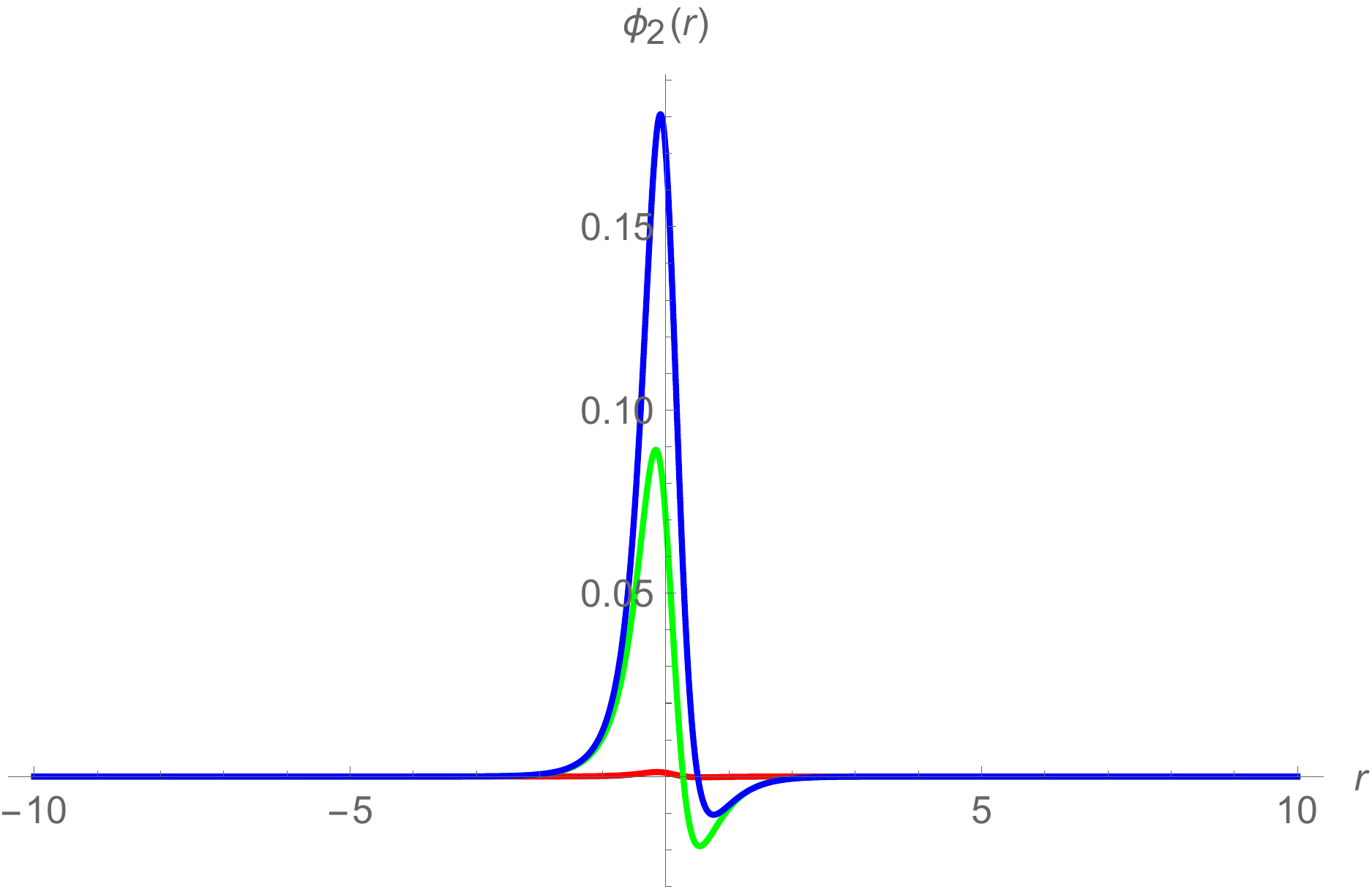}
  \caption{$\phi_2(r)$ solutions}
  \end{subfigure}
  \begin{subfigure}[b]{0.45\linewidth}
    \includegraphics[width=\linewidth]{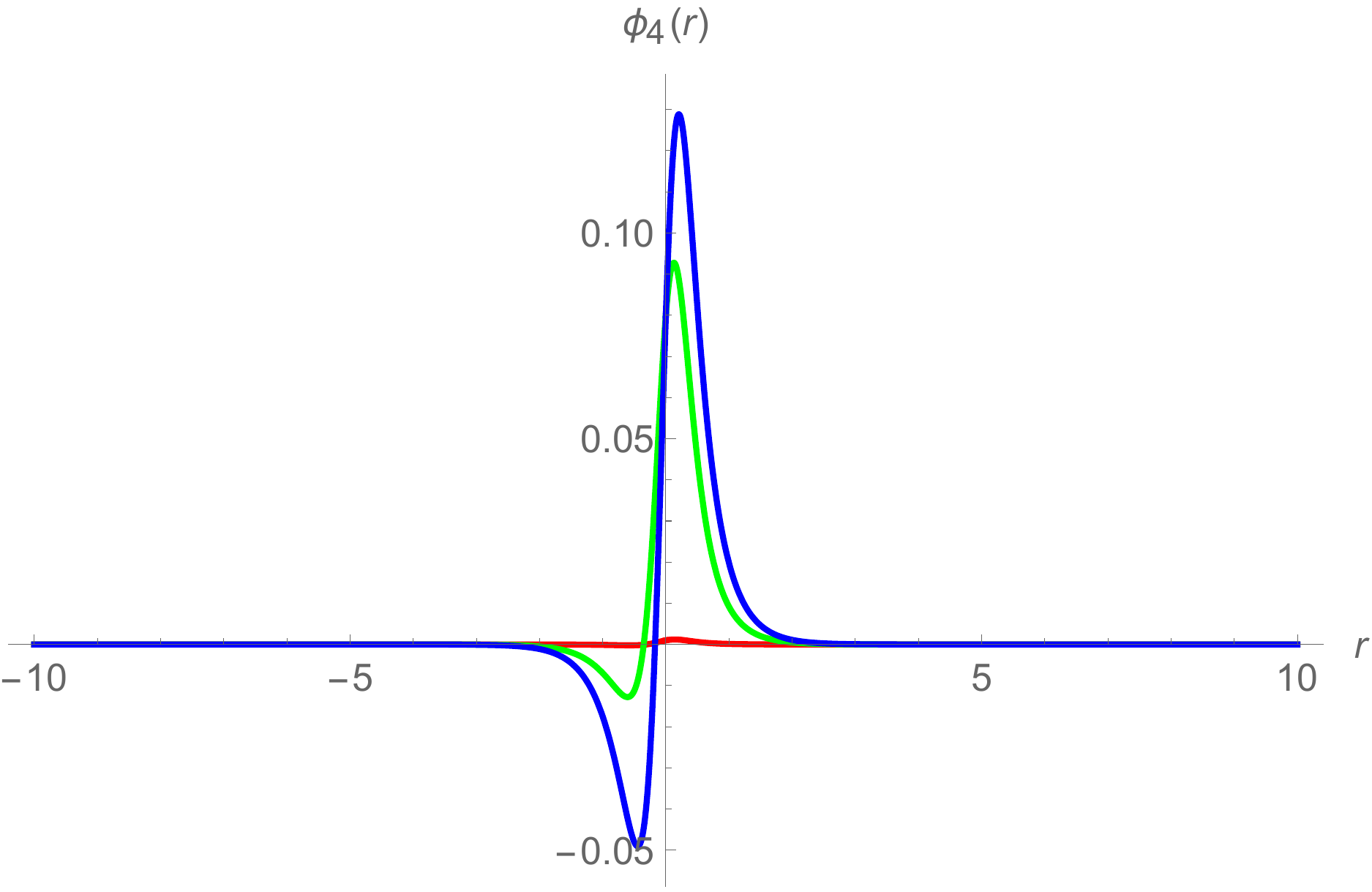}
  \caption{$\phi_4(r)$ solutions}
  \end{subfigure}\\
  \begin{subfigure}[b]{0.45\linewidth}
    \includegraphics[width=\linewidth]{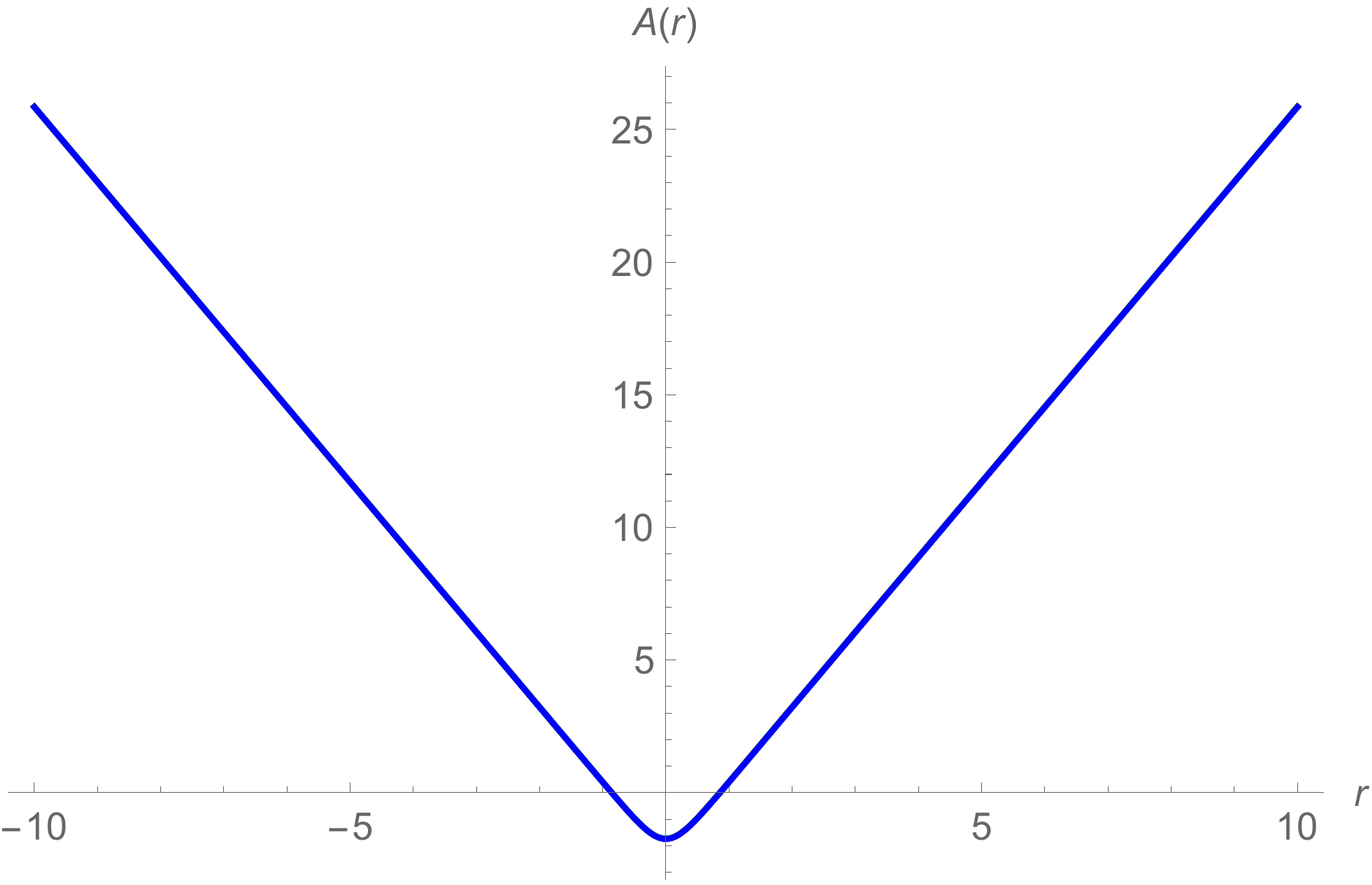}
  \caption{$A(r)$ solutions} 
  \end{subfigure}
  \caption{$N=2$ supersymmetric Janus solutions from $N=4$ $SO(4)\times SO(4)$ gauged supergravity for $g_1=g_2=2\sqrt{2}$, $\ell=2$ and $\kappa=1$ with different boundary conditions.}
  \label{FigJ2}
\end{figure}

We now look at more complicated solutions in matter-coupled $N=4$ gauged supergravity. We first note that Janus solutions are not possible for all the subtruncations with $\chi=0$. We then consider the case of $N=2$ supersymmetry and $SO(2)\times SO(2)\times SO(3)\times SO(2)$ symmetry with $\phi_1=\phi_3=0$ and $\epsilon^{2}=\epsilon^3=0$. The other equivalent cases can be obtained by replacing $(\phi_2,\phi_4)$ and/or $(\epsilon^{1},\epsilon^4)$ by $(\phi_1,\phi_3)$ and/or $(\epsilon^{2},\epsilon^3)$.
\\
\indent With the superpotential \eqref{W_N2_gen} and the same procedure as in the previous case, we find the following BPS equations
\begin{eqnarray}
\phi_2'&=&-2\textrm{sech}^2\phi_4\frac{A'}{W}\frac{\pd W}{\pd \phi_2}-2\textrm{sech}\phi_4 \frac{\kappa e^{-A}}{\ell W}\frac{\pd W}{\pd \phi_4},\\
\phi_4'&=&-\frac{A'}{W}\frac{\pd W}{\pd \phi_4}+2\textrm{sech}\phi_4 \frac{\kappa e^{-A}}{\ell W}\frac{\pd W}{\pd \phi_2},\\
\phi'&=&-4\frac{A'}{W}\frac{\pd W}{\pd \phi}-4e^\phi \frac{\kappa e^{-A}}{\ell W}\frac{\pd W}{\pd \chi},\\
\chi'&=&-4e^{2\phi}\frac{A'}{W}\frac{\pd W}{\pd \chi}+4e^\phi \frac{\kappa e^{-A}}{\ell W}\frac{\pd W}{\pd \phi},\\
A'^2&=&W^2-\frac{e^{-2A}}{\ell^2}
\end{eqnarray}
with 
\begin{eqnarray}
\frac{1}{W}\frac{\pd W}{\pd \phi_2}&=& \frac{1}{4W^2}g_1e^{-\phi}\cosh\phi_4\left[\sinh\phi_2(g_2e^\phi+g_1\cosh\phi_2\cosh\phi_4)\right. \nonumber \\
& &\left.+g_2\chi\cosh\phi_2\sinh\phi_4\right],\\
\frac{1}{W}\frac{\pd W}{\pd \phi_4}&=& \frac{1}{4W^2}\left[e^{-\phi}(g_1\sinh\phi_2\sinh\phi_4\right.\nonumber \\
& &+g_2\chi\cosh\phi_4)(g_1\cosh\phi_4\sinh\phi_2+g_2\chi\sinh\phi_4) \nonumber \\
& &\left. g_2\sinh\phi_4(g_1\cosh\phi_2+g_2e^{\phi}\cosh\phi_4)\right],\\
\frac{1}{W}\frac{\pd W}{\pd \phi}&=&\frac{1}{8W^2}\left[2g_2\cosh\phi_4(g_1\cosh\phi_2+g_2e^\phi \cosh\phi_4)\right.\nonumber \\
& &-e^{-\phi}(g_1\sinh\phi_2\sinh\phi_4+g_2\chi\cosh\phi_4)^2\nonumber \\
& &\left.-e^{-\phi}(g_1\cosh\phi_2+g_2e^\phi\cosh\phi_4)^2 \right],\\
\frac{1}{W}\frac{\pd W}{\pd \chi}&=& \frac{1}{4W^2}g_2e^{-\phi}\cosh\phi_4(g_1\sinh\phi_2\sinh\phi_4+g_2\chi\cosh\phi_4)\, .
\end{eqnarray}   
\indent Examples of numerical solutions are given in figure \ref{FigJ2}. In the figure, the solutions for the warped factor $A$ are very close to each other. In this case, the solutions describe conformal interfaces with $N=(2,0)$ or $N=(0,2)$ supersymmetries corresponding respectively to $\kappa=1$ or $\kappa=-1$ and could provide a holographic description of deformations by position-dependent operators within the dual $N=4$ SCFT. 

\section{Conclusions and discussions}\label{conclusion}
In this paper, we have studied a number of holographic RG flows from matter-coupled $N=4$ gauged supergravity with $SO(4)\times SO(4)$ gauge group by truncating to $SO(2)\times SO(2)\times SO(2)\times SO(2)$ singlet scalars. For vanishing axion, there exist $N=4$ supersymmetric flow solutions with $SO(2)\times SO(3)\times SO(2)\times SO(2)$ and $SO(2)\times SO(2)\times SO(2)\times SO(2)$ symmetries. In the presence of axion, the solutions preserve only $N=2$ supersymmetry and $SO(2)\times SO(2)\times SO(3)\times SO(2)$ symmetry. All of these solutions are singular at finite values of the radial coordinate, but all the singularities turn out to be physically acceptable according to the criterion given in \cite{Gubser_singularity}. The solutions can be interpreted as RG flows from the $N=4$ SCFT, dual to the unique $N=4$ supersymmetric $AdS_4$ vacuum within this truncation, to non-conformal phases in the IR with different global symmetries. All of these solutions could hopefully be useful in studying various mass deformations of strongly coupled $N=4$ SCFT via AdS/CFT holography. 
\\
\indent We have also found $N=4$ and $N=2$ supersymmetric Janus solutions with $SO(4)\times SO(4)$ and $SO(2)\times SO(2)\times SO(3)\times SO(2)$ symmetries. The former is obtained by truncating out all scalars from the vector multiplets resulting in the solutions of pure $N=4$ gauged supergravity while the latter is a genuine solution of the matter-coupled theory. These solutions holographically describe two-dimensional conformal interfaces in the dual $N=4$ SCFT with unbroken $N=(4,0)$ or $N=(2,0)$ supersymmetries on the interfaces. Generally, these configurations are identified with deformations of the dual SCFT by operators or vacuum expectation values that depend on the coordinate transverse to the interfaces.
\\
\indent It would be interesting to explicitly identify the dual $N=4$ SCFT together with relevant deformations dual to the solutions found here. Uplifting the $SO(4)\times SO(4)$ gauged supergravity considered here to higher dimensions will allow to embed these solutions in ten or eleven dimensions giving rise to new examples of AdS/CFT duality in the context of string/M-theory. In particular, it could be interesting to check whether the singularities allowed by the criterion of 
\cite{Gubser_singularity} are physical in string/M-theory by using the criterion of \cite{Maldacena_Nunez_nogo}. If this is indeed the case, identifying relevant M-brane or D-brane configurations related to these four-dimensional solutions also deserves further study. Finally, finding more general holographic solutions of $N=4$ gauged supergravities in other truncations or in other gauge groups could be interesting as well. These solutions might include holographic solutions describing spatially varying mass terms given in \cite{Gaunlett_spatial_mass1} and \cite{Gaunlett_spatial_mass2}. We leave all these issues for future investigation. 
\vspace{0.5cm}\\
{\large{\textbf{Acknowledgement}}} \\
This work is supported by The Thailand Research Fund (TRF) under grant RSA6280022. It is a pleasure to thank Patharadanai Nuchino for reading the manuscript.

\end{document}